\documentclass[12pt]{article}

\pdfoutput=1

\usepackage{verbatim}
\usepackage{setspace}
\usepackage{times,amssymb}
\usepackage{graphicx}
\usepackage{microtype}
\usepackage{siunitx}
\usepackage{booktabs}
\usepackage[intlimits]{amsmath}
\usepackage{blindtext}
\usepackage{enumitem}
\usepackage{cite}
\usepackage[font=small]{caption}
\usepackage[usenames, dvipsnames]{color}

\usepackage{soul}

\usepackage{titling}

\usepackage{ragged2e}

\usepackage{hyperref}
\usepackage{physics}
\usepackage{epstopdf}
\usepackage{float}
\usepackage{amsfonts}
\usepackage{amsmath,amssymb}
\newcommand{\eqn}[1]{{Eq.~(\ref{#1})}}
\newcommand{\fig}[1]{{Fig.~(\ref{#1})}}

\title{Coherent vortex dynamics in a strongly-interacting superfluid on a silicon chip} 

\author
{Yauhen P. Sachkou,$^{1\ast}$ Christopher G. Baker,$^{1\ast}$ Glen I. Harris,$^{1\ast}$, \\
Oliver R. Stockdale,$^{2}$  Stefan Forstner,$^{1}$ Matthew T. Reeves,$^{2}$ Xin He,$^{1}$  \\
David L. McAuslan,$^{1}$ Ashton S. Bradley,$^{3}$ Matthew J. Davis,$^{1,2}$ \& \\
Warwick P. Bowen$^{1\dagger}$\\
\\
\normalsize{$^{1}$ARC Centre of Excellence for Engineered Quantum Systems, School of Mathematics and} \\ 
 \normalsize{Physics, University of Queensland, St Lucia, QLD 4072, Australia.} \\ 
\normalsize{$^{2}$ARC Centre of Excellence in Future Low-Energy Electronics Technologies,}\\
\normalsize{School of Mathematics and Physics, University of Queensland, St Lucia, QLD 4072, Australia.} \\
\normalsize{$^{3}$Department of Physics, Centre for Quantum Science, and Dodd-Walls Centre for Photonic}\\
\normalsize{and Quantum Technologies, University of Otago, Dunedin, New Zealand.}\\
\\
\normalsize{$^\ast$These authors contributed equally to this work.}\\
\normalsize{$^\dagger$To whom correspondence should be addressed; E-mail: w.bowen@uq.edu.au.}
}

\date{}

\begin{document} 

\baselineskip24pt

\maketitle

\vspace{-5 mm}

\begin{abstract}

\normalsize

\noindent
Two-dimensional superfluidity and quantum turbulence are directly connected to the microscopic dynamics of quantized vortices. However, surface effects have prevented direct observations of  coherent vortex dynamics in strongly-interacting two-dimensional systems. Here, we overcome this challenge by confining a two-dimensional droplet of superfluid helium at microscale on the atomically-smooth surface of a silicon chip. An on-chip optical microcavity allows laser-initiation of vortex clusters and nondestructive observation  of their  decay in a single shot. Coherent dynamics dominate, with thermal vortex diffusion suppressed by six orders-of-magnitude. This establishes a new on-chip platform to study emergent phenomena in strongly-interacting superfluids, test astrophysical dynamics such as those in the superfluid core of neutron stars in the laboratory, and construct quantum technologies such as precision inertial sensors.

\end{abstract}

\onehalfspacing

Strongly-interacting many-body quantum systems exhibit rich behaviours of significance to areas ranging from superconductivity~\cite{mitra_quantum_2018} to quantum computation~\cite{gross_quantum_2017,king_observation_2018},  astrophysics~\cite{page_rapid_2011,chamel_superfluidity_2017,the_star_collaboration_global_2017}, and even string theory~\cite{kovtun_viscosity_2005}. The first example of such a behaviour, superfluidity, was discovered more than eighty years ago in cryogenically cooled liquid helium-4 \cite{kapitza_viscosity_1938}. Quite remarkably, it was found to persist even in thin two-dimensional  films~\cite{atkins_third_1959}, for which the well-known Mermin-Wagner theorem precludes condensation into a superfluid phase in the thermodynamic limit~\cite{mermin_absence_1966}. This apparent contradiction was resolved by Berezinskii, Kosterlitz and Thouless (BKT), who predicted that quantized vortices allow a topological phase transition into superfluidity\cite{kosterlitz_long_1972,bishop_study_1978}. It is now recognized that quantized vortices also dominate much of the out-of-equilibrium dynamics of two-dimensional superfluids, such as quantum turbulence\cite{neely_characteristics_2013}.

Recently, laser control and imaging of vortices in ultracold gases~\cite{freilich_real-time_2010,donadello_observation_2014} and semiconductor exciton-polariton systems~\cite{amo_polariton_2011,estrecho_single-shot_2018} has provided rich capabilities to study superfluid dynamics~\cite{navon_critical_2015} including, for example, the formation of collective vortex dipoles with negative temperature and large-scale order~\cite{gauthier_negative-temperature_2018,johnstone_order_2018} as predicted by Lars Onsager seventy years ago~\cite{onsager_statistical_1949}. However, these experiments are generally limited to the regime of weak interactions, where the Gross-Pitaevskii equation provides a microscopic model of the dynamics of the  superfluid. The regime of strong interactions can be reached  by tuning the atomic scattering length in ultracold gases~\cite{makotyn_universal_2014,zwierlein_vortices_2005}. However, technical challenges have limited investigations of nonequilibrium phenomena~\cite{makotyn_universal_2014}. The strongly-interacting regime defies a microscopic theoretical treatment and is the relevant regime for superfluid helium as well as for astrophysical superfluid phenomena such as pulsar glitches~\cite{anderson_pulsar_1975} and superfluidity of the quark-gluon plasma in the early universe~\cite{the_star_collaboration_global_2017}. The vortex dynamics in this regime are typically predicted using phenomenological  vortex models. However, whether the vortices should have inertia~\cite{thouless_vortex_2007,simula_vortex_2018}, the precise nature of the forces they experience due to the normal component of the fluid~\cite{sonin_magnus_1997}, and how to treat dissipation given the non-local nature of the vortex flow fields~\cite{adams_vortex_1987,thompson_quantum_2012} all remain unclear. Moreover, point-vortex modelling offers limited insight into the process of vortex creation and annihilation, which are crucial to understand the dynamics of topological phase transitions. 

Here, we report the observation of coherent vortex dynamics in a strongly-interacting two-dimensional superfluid. We achieve this by developing a microscale photonic platform to initialize vortex clusters in two-dimensional helium-4 on a silicon chip, confine them, and image their spatial distribution over time. Our experiments characterize  vortex distributions via their interactions with resonant sound waves, leveraging ultraprecise sensing methods from cavity optomechanics~\cite{purdy_quantum_2017,basiri-esfahani_precision_2019,harris_laser_2016,kashkanova_superfluid_2017}. Microscale confinement greatly enhances the vortex-sound interactions, and enables resolution of the dynamics of few-vortex clusters in a single-shot and tracked over many minutes, as they interact, dissipate energy and annihilate. We observe evaporative heating where the annihilation of low-energy vortices causes an increase in the kinetic energy of the remaining free vortices. Strikingly, these annihilation events draw energy out of a background flow, causing a net increase in free-vortex kinetic energy as the system evolves. 

Our experiments verify a thirty-year-old prediction that the vortex diffusivity should become exceptionally small when operating at temperatures far below the BKT transition~\cite{adams_vortex_1987}, yielding a diffusivity six orders-of-magnitude lower than has been observed  previously for unpinned vortices in superfluid helium films~\cite{adams_vortex_1987}. As a consequence, the system operates well within the regime of coherent vortex dynamics, with the timescale for dissipation found to exceed the coherent evolution time by more than five orders-of-magnitude. The on-chip platform reported here provides a new technology to explore the dynamics of phase transitions and quantum turbulence in strongly-interacting superfluids, and to study how such fluids evolve towards thermal equilibrium and how they dissipate energy. It may also allow new phenomena to be engineered through strong sound-vortex interactions, and the development of superfluid matter-wave circuitry on a silicon chip.

\section{Interactions with sound allow vortex imaging}

\begin{figure}[t!]
\centering
\includegraphics[width = 0.6\textwidth]{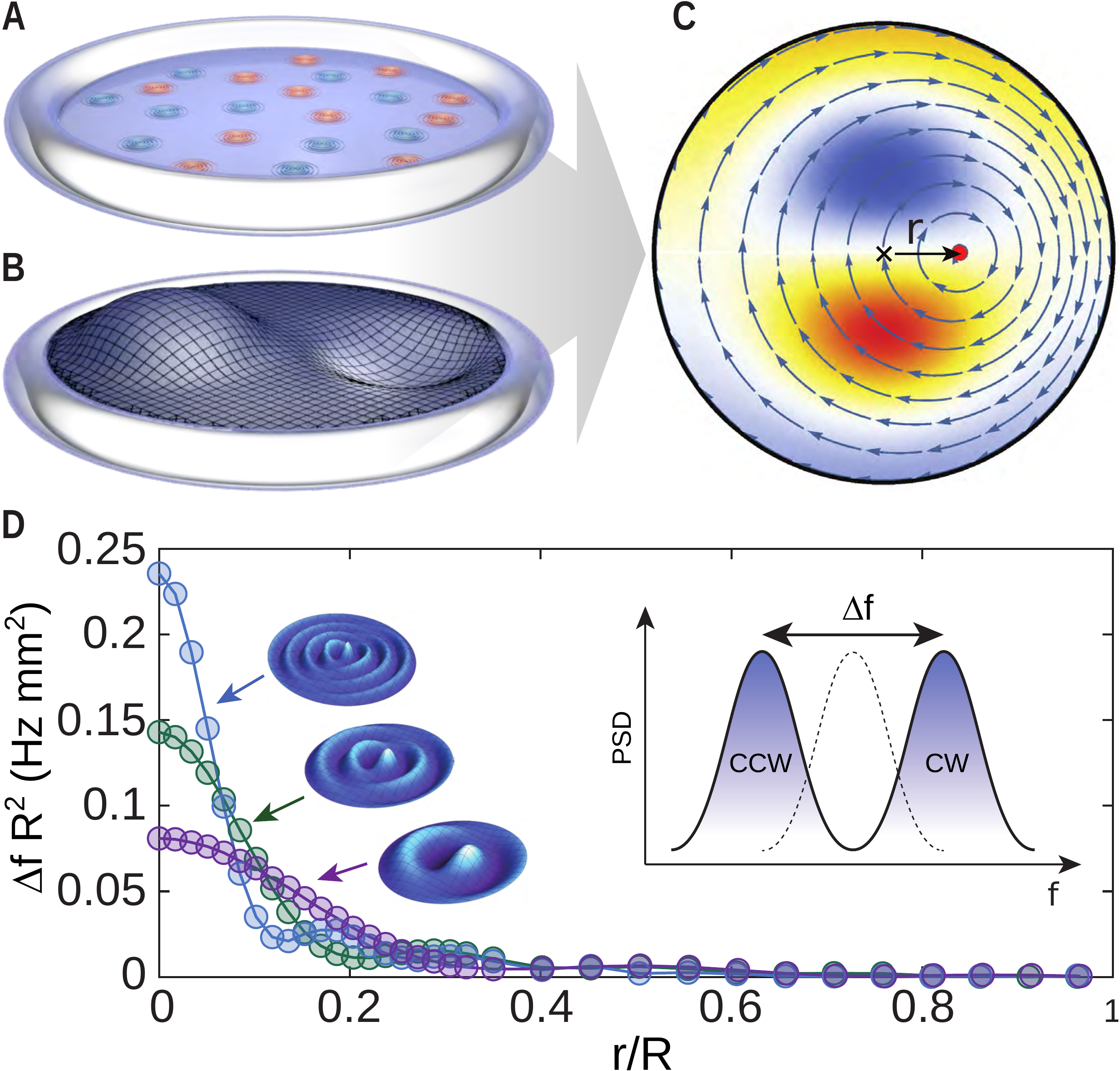} 
\caption{{\bf Vortex-sound interactions on a disk.} Confinement of both vortices \textbf{(A)} and sound \textbf{(B)} within the same microscale domain enhances the vortex-sound interaction rate. Vortices and sound couple via their flow fields, as shown in \textbf{(C)}. Surface color: sound mode amplitude profile; blue lines: vortex streamlines. The red dot represents a vortex offset from the disk origin by distance $r$. \textbf{(D)} Normalized frequency splitting per vortex for sound modes $(m,n) = (1,3)$, $(1,5)$ and $(1,8)$ calculated via finite-element modelling~\cite{forstner_modelling_2019}, with their respective spatial profiles. The inset schematically depicts the vortex-induced splitting $\Delta f$ between clockwise and counter-clockwise sound modes in the presence of a CW vortex.}
\label{Fig1main}
\end{figure}

The interaction between light and vortices is extremely weak in two-dimensional helium due to the few-nanometer film thickness,  \si{\angstrom}ngstr\"{o}m-scale of vortex cores, and the exceedingly low refractive index of superfluid helium. This precludes direct optical imaging techniques similar to those used to image vortex dynamics in bulk three-dimensional helium~\cite{fonda_direct_2014}. Instead, in our experiments the vortex dynamics are tracked via their influence on sound waves. The vortices and sound waves co-exist in a thin superfluid helium film. They are geometrically confined on the bottom surface of a microtoroidal cavity which supports optical whispering gallery modes and is held above a silicon substrate on a pedestal. Figure~\ref{Fig1main} illustrates the interactions between vortices (Fig.~\ref{Fig1main}A) and sound modes (Fig.~\ref{Fig1main}B) confined to the surface of a disk. The sound modes are third-sound~\cite{atkins_third_1959} -- surface waves analogous to shallow-water waves but with a restoring force provided by the van der Waals interaction with the substrate. They are well described by resonant Bessel modes of the first kind, which are characterized by their radial $m$ and azimuthal $n$ mode numbers~\cite{baker_theoretical_2016}.
 
The vortex flow field causes Doppler shifts of the frequencies of the sound modes, lifting the degeneracy between clockwise and counter-clockwise waves (Fig.~\ref{Fig1main}C\&D) \cite{ellis_quantum_1993,forstner_modelling_2019}. The magnitude of the frequency shift induced by a vortex depends both on its position and on the spatial profile of the sound mode, as shown for several modes in Fig.~\ref{Fig1main}D.  Despite the strong interactions between helium atoms, the phase coherence and incompressibility of the superfluid combine to ensure linearity in our experiments. As such, the total flow field of a vortex cluster is given by the linear superposition of the flow of each constituent vortex. The total splitting between counter-rotating sound modes is then equal to the sum of the splittings generated by each vortex. We exploit this linearity, combined with the vortex-position dependent interaction and simultaneous measurements of splitting on several sound modes, to characterize the spatial distribution of vortex clusters in a manner analogous to experiments that use multiple cantilever eigenmodes to image the distribution of deposited nanoparticles \cite{hanay_inertial_2015}.

\section{Nonequilibrium vortex clusters can be generated with laser light}

\begin{figure}[t!]
\centering
\includegraphics[width = \textwidth]{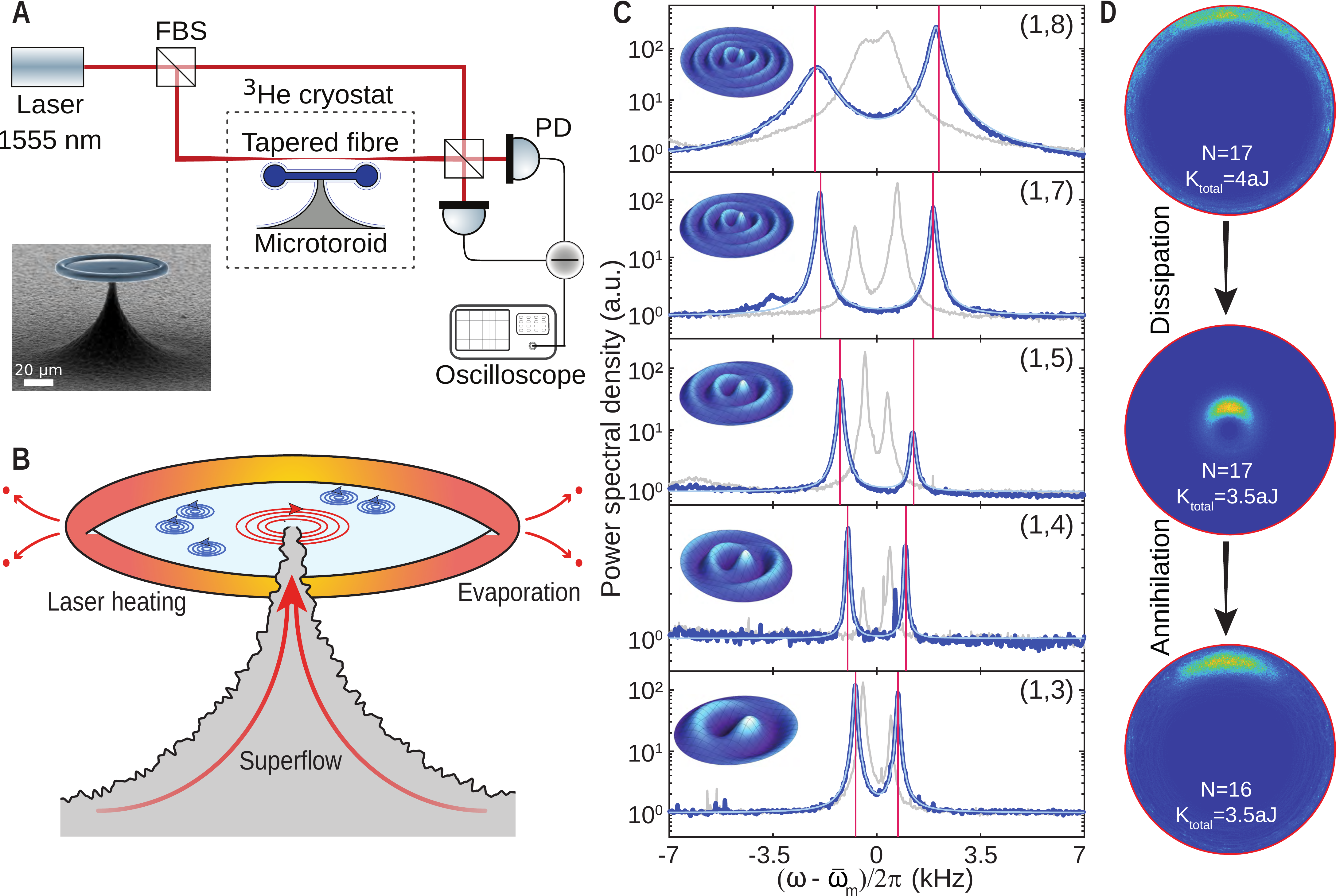} 
\caption{{\bf Laser initialization and observation of vortex clusters.} \textbf{(A)} Experimental set-up: balanced homodyne detection scheme implemented within a  fiber interferometer. FBS: fiber beam splitter, PD: photodetector. Inset: scanning electron microscope image of the microtoroidal optical cavity used in the experiments. Scale bar: 20 $\mu$m. \textbf{(B)} Sketch of the vortex generation process. \textbf{(C)} Splittings of $(m,n) = (1,3)$, $(1,4)$, $(1,5)$, $(1,7)$ and $(1,8)$ third-sound modes immediately after the initialization process. Spatial amplitude profiles of the split modes are shown as insets. Red bars mark the theoretically computed splittings right after the system initialization. Grey spectra correspond to the unperturbed third-sound modes without vortex ensemble initialization. The residual splitting in these unperturbed spectra is due to irregularities in the circularity of the microtoroid that break the degeneracy between standing-wave Bessel modes\cite{harris_laser_2016}. We accounted for these in data processing [section~\ref{subsec:geo_splitting} of Supplementary Materials]. \textbf{(D)} Three exemplar metastable distributions showing the effects of dissipation and annihilation. The colormap indicates the free-vortex probability density.}
\label{Fig2main}
\end{figure}

Figure~\ref{Fig2main}A shows a schematic of the experiment. The microtoroid is placed inside a sealed sample chamber within a closed-cycle $^3$He cryostat and optically excited via an optical nanofiber. The sample chamber is filled with $^4$He gas which, upon cooling across the superfluid critical temperature, condenses into a superfluid film of thickness $d\sim7.5$~nm, coating the inside of the chamber including the optical microcavity \cite{harris_laser_2016,mcauslan_microphotonic_2016}. Motion of the superfluid film manifests as fluctuations of the phase of light confined inside the cavity, which are resolved via balanced homodyne detection implemented within a fiber interferometer \cite{harris_laser_2016}. Frequency analysis of the output photocurrent reveals third-sound modes, with resonance frequencies in a good agreement with the expected Bessel mode frequencies (Fig.~\ref{Fig2main}A) [see sections~\ref{Sec:Experimental_details}~$\&$~\ref{Sec:modes_identification} of Supplementary Materials]. 

We find experimentally that vortex clusters can be optically initialized into a vortex dipole in several ways, including pulsing the intensity of the injected laser to induce superfluid flow via the `\textit{fountain effect}' \cite{tilley_superfluidity_1990}, and optomechanical driving of low-frequency third-sound modes via dynamical backaction \cite{harris_laser_2016}. Both of these techniques induce  flow that exceeds the superfluid critical velocity\cite{tilley_superfluidity_1990}, triggering the generation of vortex pairs (Fig.~\ref{Fig2main}B). In the presence of a circular boundary, an ensemble of vortex pairs evolves into a metastable state characterized at high energies by a large-scale negative-temperature Onsager vortex dipole~\cite{onsager_statistical_1949,gauthier_negative-temperature_2018}. In our case, the microtoroid pedestal introduces a deep potential that vortices can pin to [section~\ref{Sec:pinning_potential} of Supplementary Materials], qualitatively modifying the physics. Vortices of one sign become pinned, creating a macroscopic circulation, while the others evolve into a free-orbiting metastable cluster [section~\ref{subsec:metastable_state} of Supplementary Materials]. As is the case  without a pinning site, this metastable state is a negative-temperature vortex dipole. We will see, however, that its dynamics displays qualitatively different features.

The timescale within which a  metastable state is reached can be estimated from the characteristic turnover time for internal rearrangement of the free-vortex cluster, $\tau \sim r_c^2 / N \kappa$, where $N$ is the number of free vortices,  $r_c$ is the radius of the cluster, and $\kappa=h/m_{\rm He}$ is the circulation quantum, with $m_{\mathrm{He}}$ being the mass of a helium atom and $h$ Planck's constant~\cite{reeves_enstrophy_2017} [section~\ref{subsec:metastable_state} of Supplementary Materials]. Taking the case of two free vortices separated by the disk radius $R=30 \, \mu \mathrm{m}$ provides an upper bound  to the turnover time of $\tau \lesssim 5$~ms. This is substantially faster than both the dissipation of the system and the temporal resolution of our measurements.
Consequently, the vortex cluster can be well approximated to exist in a metastable state throughout its evolution, with this state modified continuously by dissipation and in discrete steps by vortex annihilation events.

Each possible metastable state is uniquely characterized by the number of free vortices, kinetic energy and angular momentum. It has been shown previously that stirring superfluid helium in an annulus generates a ring of free vortices~\cite{fetter_low-lying_1967}. Assuming our initialization process produces such a ring allows the parameter space to be reduced to vortex number and kinetic energy. Performing point-vortex simulations for vortex configurations initially arranged in a ring, and varying both the ring radii and vortex number, we determine the possible metastable vortex distributions as a function of these two parameters [section~\ref{subsec:metastable_state} of Supplementary Materials].  In their metastable state, the free vortices exist in an orbiting horse-shoe shaped cluster separated from the origin, as illustrated in Fig.~\ref{Fig2main}D. Together with the macroscopic circulation, this forms a vortex dipole with separation that shrinks as the system dissipates kinetic energy and grows when vortices are annihilated (see Fig.~\ref{Fig2main}D).

\section{Vortex clusters evolve coherently}

To determine the instantaneous metastable vortex distribution from a single continuous measurement, we generate a nonequilibrium vortex cluster by optically initiating supercritical flow. We then simultaneously measure the frequency shifts induced on several third-sound modes (Fig.~\ref{Fig2main}C) as the cluster  evolves over time. 
 Using the vortex-position dependent splitting function $\Delta f(r)$ (see Fig.~\ref{Fig1main}D), the frequency shifts expected on the $(m,n)=(1,3)$, $(1,4)$, $(1,5)$, $(1,7)$, $(1,8)$ third-sound modes are computed for each possible metastable distribution and compared to the observed shifts. This allows us to ascertain both the metastable state that most closely matches the observed frequency shifts at a given time, as well as the range of metastable states for which the shifts are statistically indistinguishable.
 
Figure~\ref{Fig2main}C shows the observed frequency shifts at the start of the measurement run, just after the vortex cluster has been initialized. As shown by the vertical red bars, we find an excellent agreement between the observed frequency splittings and those obtained with the optimal metastable distribution. Moreover, we find that the statistical uncertainty in the vortex number and kinetic energy is relatively small. We identify the initial metastable vortex distribution as a vortex dipole either containing 17 free vortices with a total kinetic energy of $K_{\mathrm{total}} = 7.8 \substack{+0.6 \\ -0.3} \, \mathrm{aJ}$, or 16 vortices with $K_{\mathrm{total}} = 8.3 \substack{+0.2 \\ -0.6} \, \mathrm{aJ}$. We are also able to determine the dipole separation, which in the case of 17 vortices is found to be $7.1 \substack{+1.3 \\ -0.7} \, \mathrm{\mu m}$
[see section~\ref{Sec:vortex_decay_models} of Supplementary Materials].

\begin{figure}[t!]
\centering
\includegraphics[width = 0.7\textwidth]{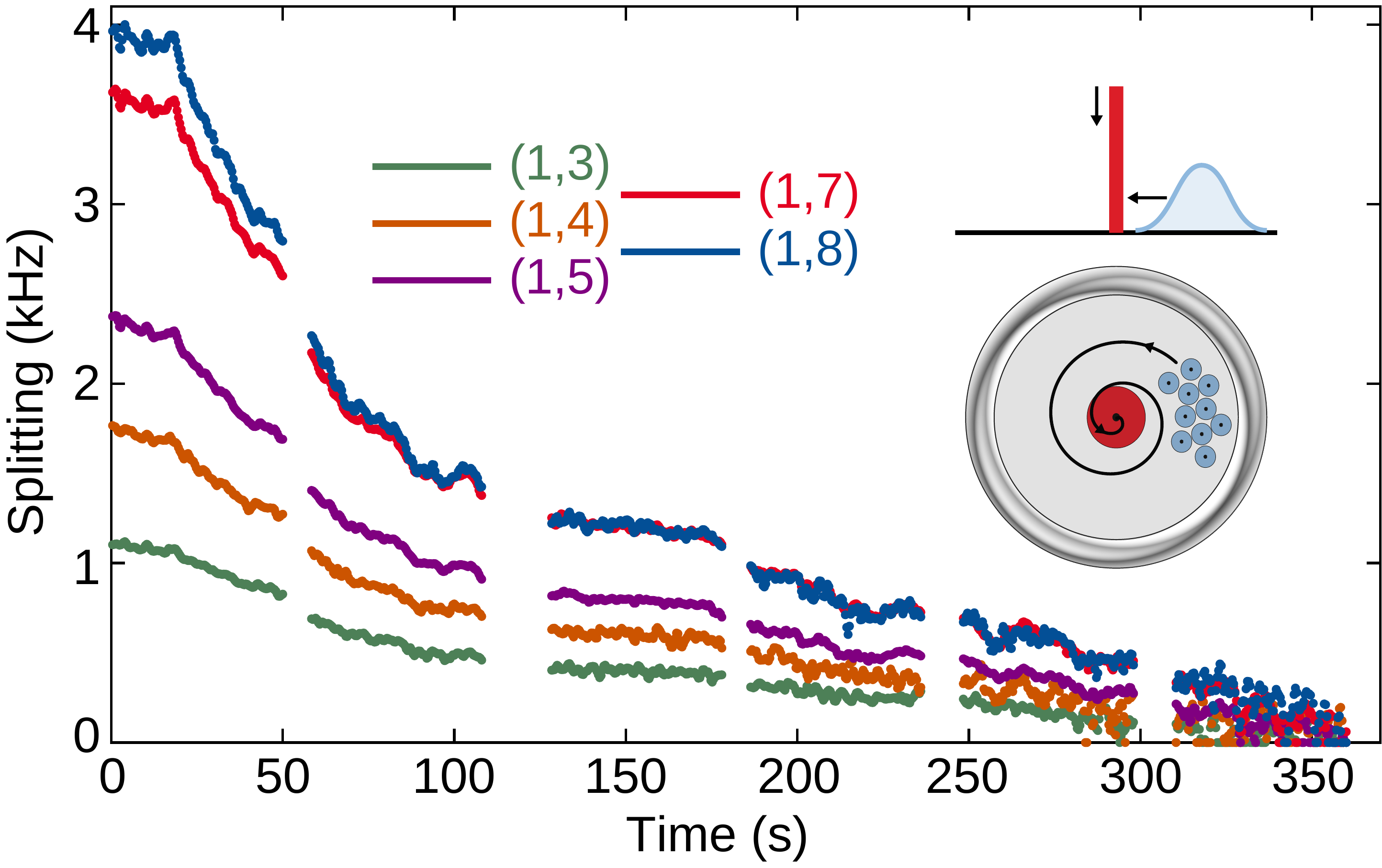}
\caption{{\bf Temporal dynamics of vortex-induced symmetry-breaking between sound waves.} Experimental splitting decay of the $(m,n)=(1,8)$, $(1,7)$, $(1,5)$, $(1,4)$ and $(1,3)$ third-sound modes (top-to-bottom traces respectively) caused by the decay of the vortex dipole. The raw data was recorded on a high bandwidth, high memory-depth oscilloscope. In this run, six continuous measurements were taken, separated by data-saving periods each of approximately ten seconds. Inset: sketch of the decay process. Red: quantized circulation around the pedestal reduces due to annihilation events. Light blue: orbiting free-vortex cluster spirals towards the origin due to dissipation. Note that the dissipation is  exaggerated in this sketch for clarity.}
\label{Fig3main}
\end{figure}

Continuously monitoring the superfluid sound modes reveals that their splitting decays over a time-scale on the order of minutes (Fig.~\ref{Fig3main}). The observed splittings are well characterized by a metastable vortex dipole throughout this entire decay process. By contrast,  they are inconsistent with other vortex dynamics models such as expansion of a  single-sign vortex cluster due to either vortex-vortex interactions or diffusive hopping between pinning sites [section~\ref{Sec:vortex_decay_models} of Supplementary Materials]. 

The kinetic energy and free-vortex number of the metastable state are shown as a function of time for a single-shot in Fig.~\ref{Fig4main}A\&B (blue curves). As one might expect, the total kinetic energy of the dipole decays continuously with time  over a period of around a minute. This decay time is comparable to previous non-spatially resolved measurements of the decay of  a persistent current~\cite{ekholm_behavior_1979}. In our experiments, the decay is accompanied by  a reduction in  the number of free vortices, as vortex-vortex interactions and dissipation drive vortices into the centre of the disk where they can annihilate with quanta of circulation of opposite sign pinned to the pedestal. These dynamics are  supported by point-vortex simulations (Fig.~\ref{Fig4main}C\&D), which show good quantitative agreement with only the dimensionless dissipation coefficient $\gamma$ as a fitting parameter. The dissipation coefficient quantifies the ratio of coherent to  dissipative timescales in the superfluid dynamics, and was found to be $\gamma \sim 2 \times 10^{-6}$ [section~\ref{subsec:implementation_of_PVM} of Supplementary Materials]. It has generally been thought that fast dissipative processes would preclude the observation of coherent dynamics in superfluid helium films. However, our experiments show that this is not the case in general, with coherent dynamics dominating by more than five orders-of-magnitude. Indeed, the dissipation coefficient obtained here is  competitive with the best ultracold atom experiments, which typically achieve $\gamma \sim 5 \times 10^{-4}$~\cite{gauthier_negative-temperature_2018,bradley_energy_2012}. The agreement between experiment and theory indicates that, within experimental uncertainties, the vortex dynamics are consistent with a simple point vortex model including local phenomenological dissipation, and without the need to introduce inertia to the vortex cores\cite{thouless_vortex_2007,simula_vortex_2018} or an Iordanskii force between vortices and the normal component of the fluid~\cite{sonin_magnus_1997}.

\section{Nonequilibrium vortex dynamics is observed in a single-shot}

\begin{figure}[t!]
\centering
\includegraphics[width = 0.9\textwidth]{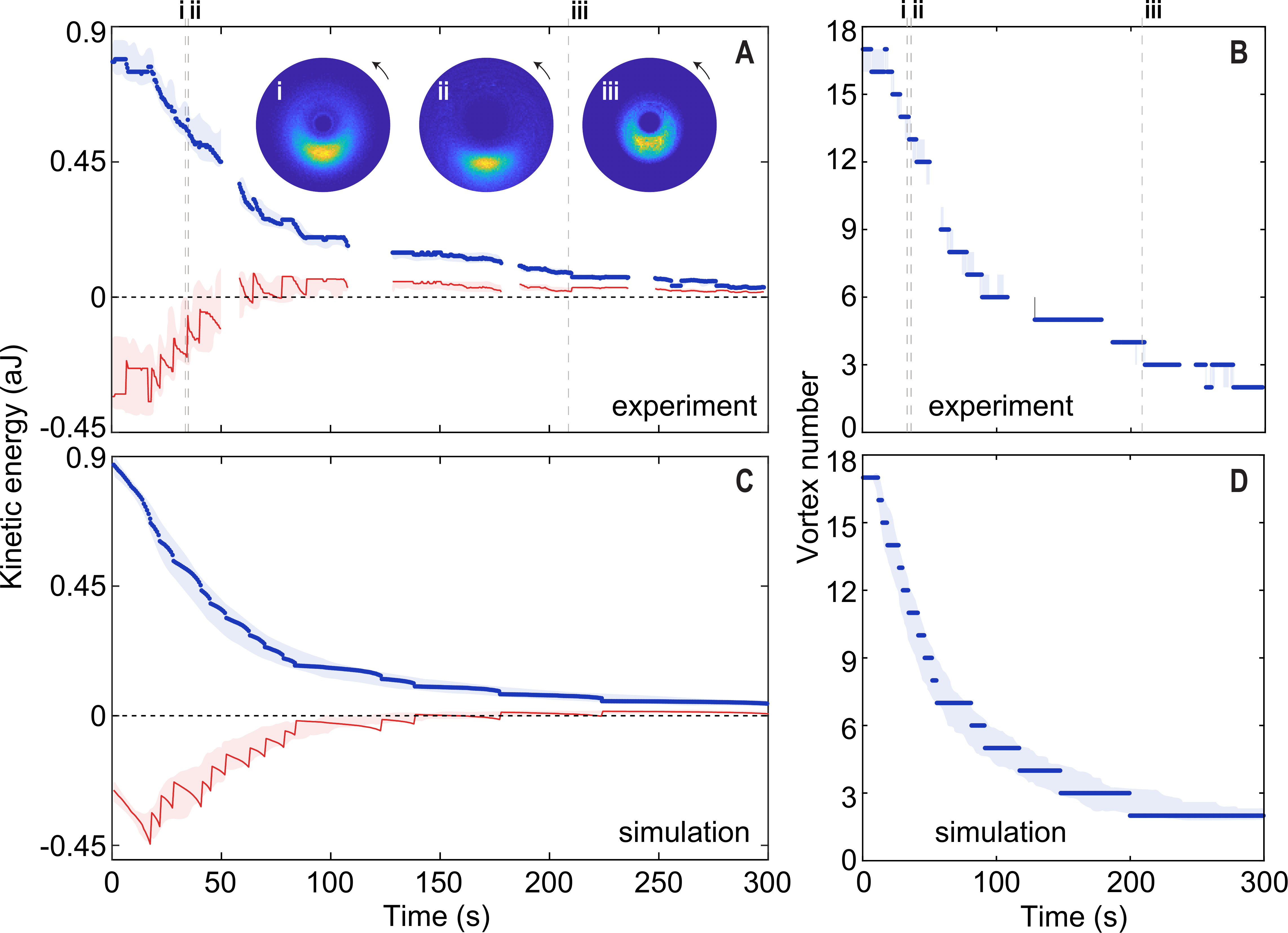} 
\caption{{\bf Single-shot evolution of vortex-cluster metastable states.} \textbf{(A)} Decay of the total kinetic energy $K_{\mathrm{total}}$ (blue curve) and increase of the kinetic energy of the free-vortex cluster $K_{\mathrm{free}}$ (red curve). Insets: quasi-equilibrium vortex distributions at times indicated by the vertical dashed lines. The first two quasi-equilibrium distributions are taken, respectively, just before and just after the 13-to-12 annihilation event. \textbf{(B)} Experimentally determined decay of the vortex number. Vertical dashed lines correspond to times in \textbf{(A)}. Note that, while this data displays steps in the vortex number, this is a feature of our analysis which minimizes the root-mean-square uncertainty only over discrete vortex number. While our experiments approach single vortex resolution, the continuous variation of vortex-induced splitting with time precludes direct unambiguous observation of individual steps in the splitting due to the creation or annihilation of vortices\cite{forstner_modelling_2019}.
 \textbf{(C)} $K_{\mathrm{total}}$ decay (blue curve) and $K_{\mathrm{free}}$ increase (red curve) extracted  from the point-vortex simulation, showing very good agreement with the experimental results in \textbf{(A)}.  \textbf{(D)} Point-vortex simulation of the vortex number decay. In all traces the shaded area corresponds to a one-standard-deviation uncertainty.}
\label{Fig4main}
\end{figure}

The pinning of vortices on the microtoroid pedestal results in a macroscopic circulation, as discussed above. The kinetic energy associated with the free vortices is given by $K_{\mathrm{free}} = K_{\mathrm{total}} - K_{\mathrm{pinned}}$, where $K_{\mathrm{pinned}} = \displaystyle{\rho d (N \kappa)^2 \ln\left(R / r_p \right) / 4 \pi}$ is the kinetic energy of the macroscopic circulation alone, with $\rho = 145$~kg/m$^3$ being the density of superfluid helium and $r_p \sim 1 \, \mu$m the radius of the pedestal. It is shown for both our experimental data and simulations by the red curves in Fig.~\ref{Fig4main}A\&C. During the first minute of evolution it is negative and, notably, increases with time. The dynamics are characterized by steps up in energy during vortex annihilation events, interspersed with a continuous dissipative decay. 

The  negativity of the free-vortex energy can be understood by considering the interference between the flow fields of a free-vortex and the macroscopic circulation. While the high flow velocity near the core of a free-vortex introduces kinetic energy, the vortex flow field also cancels a component of  the background flow. For a sufficiently large circulation, this cancellation effect dominates, leading to an overall negative energy cost to introducing the vortex cluster [section~\ref{subsec:metastability_of_pinned_flow_around_pedestal} of Supplementary Materials].
  
The increase in free-vortex energy over time can be explained by considering the process of vortex annihilation in a macroscopic background flow. To annihilate, a free vortex must reach the pedestal, where its contribution to the total kinetic energy is at a minimum. To do this, it gives up kinetic energy to the remaining free vortices. This process of removing low energy vortices has been described as {\it evaporative heating}~\cite{simula_emergence_2014}, in analogy to evaporative cooling of ultracold atomic ensembles~\cite{anderson_observation_1995}. However, while standard evaporative heating can explain a {\it per-vortex} increase in kinetic energy\cite{johnstone_order_2018}, its effect is to reduce the {\it net} free-vortex kinetic energy\cite{johnstone_order_2018}. The physics is modified here by the presence of a macroscopic background flow. The annihilation of a free vortex with a pinned circulation quanta cancels a component of this flow, reducing its kinetic energy while leaving the total kinetic energy essentially unchanged, as can be seen by the lack of discrete steps in the blue curves of Fig.~\ref{Fig4main}A\&C. We see, therefore, that annihilation events increase the kinetic energy of the free-vortex cluster by drawing energy out  of the background flow. This pushes the cluster outwards to a higher separation, as illustrated by the metastable states just prior to, and after, an annihilation event in insets $i$ and $ii$ of  Fig.~\ref{Fig4main}A. At later times, dissipation causes the separation to decay as shown in inset $iii$. 
   
As well as allowing the observation of evaporative heating in a single continuous shot, our experiments allow the diffusivity $D$ of vortices to be established in a new regime for two-dimensional superfluid helium.
Substantial research efforts have been devoted to understanding vortex diffusivity via its effect on the damping and resonance frequency of a mechanical resonator~\cite{adams_vortex_1987} and on the attenuation of third-sound~\cite{kim_vortex_1984}, particularly in the context of dynamic corrections to the BKT transition. However, those experiments could only interrogate the fluid at temperatures within a few tens of millikelvin of the BKT transition, where the diffusivity is large enough to provide an unambiguous signal. In contrast, our experiments allow the diffusivity to be characterized at a superfluid temperature $T\sim 500$~mK, far below the BKT transition temperature of around $1$~K, where the superfluid fraction exponentially approaches unity. We obtain a value of $D=k_B T \gamma/\rho d \kappa \sim 100$~nm$^2\,$s$^{-1}$, where $T \sim 500$~mK is the superfluid film temperature and $k_B$ is the Boltzmann constant. This diffusivity is six orders-of-magnitude below previous measurements where the vortex dynamics is not dominated by pinning~\cite{adams_vortex_1987}, and verifies predictions made over thirty years ago that 
the diffusivity should be exceedingly small at low temperatures~\cite{adams_vortex_1987}.

\section{Concluding perspectives}

The experiments reported here were enabled by the greatly enhanced vortex-vortex, vortex-sound, and sound-light interactions provided by microscale confinement. Interactions with strong pinning sites have previously prevented the observation of coherent vortex dynamics in two-dimensional superfluid helium~\cite{ellis_quantum_1993}. In our experiments, vortex-vortex interactions dominate due to the increased confinement and atomically smooth surface of the device. The smoothness of the surface results in a conservative upper bound to the vortex unpinning velocity of 0.2~cm/s [section~\ref{Sec:vortex_pinning_on_surfaces} of Supplementary Materials], three orders-of-magnitude lower than previous experiments~\cite{ellis_quantum_1993}. As such, vortex-vortex interactions dominate even for the smallest possible clusters containing only two vortices. Furthermore, a four-order-of-magnitude enhancement in vortex-sound interactions compared to earlier experiments~\cite{ellis_observation_1989,ellis_quantum_1993} allows resolution approaching the single-vortex level. Together, these capabilities provide a new tool for future study of the rich dynamics of strongly-interacting two-dimensional superfluids on a silicon chip. 
 
The ability to nondestructively track vortex dynamics in a single shot opens the prospect to explore out-of-equilibrium dynamics and stochastic noise-driven processes that are challenging to study with other techniques~\cite{serafini_vortex_2017,estrecho_single-shot_2018}. It also promises to resolve contentious aspects of vortex dynamics in strongly-interacting superfluids, such as dissipation/diffusion models \cite{adams_vortex_1987,thompson_quantum_2012}, vortex inertia \cite{thouless_vortex_2007,simula_vortex_2018} and the Iordanskii force \cite{sonin_magnus_1997}. Furthermore, while the experiments reported here were performed with a relatively small number of vortices, the  \si{\angstrom}ngstr\"{o}m-scale of the vortex core in helium-4 will enable the future  research on the dynamics of ensembles of thousands of vortices, a regime well outside current capabilities with cold atom and exciton-polariton superfluids~\cite{reeves_enstrophy_2017}. This could allow emergent phenomena in two-dimensional turbulence to be explored, such as the inverse energy cascade~\cite{rutgers_forced_1998,reeves_enstrophy_2017} and anomalous hydrodynamics~\cite{wiegmann_anomalous_2014,yu_emergent_2017}.

\section*{Acknowledgements}
This work was funded by the U.S. Army Research Office through grant number W911NF17-1-0310. It was also partially supported by the Australian Research Council Centre of Excellence for Engineered Quantum Systems (EQUS, Project No. CE170100009), and the Australian Research Council Centre of Excellence in Future Low-Energy Electronics Technologies (FLEET, Project No. CE170100039). W.P.B. acknowledges the Australian Research Council Future Fellowship FT140100650. C.G.B acknowledges a Fellowship from the University of Queensland (UQFEL1833877). The authors acknowledge insightful discussions with Xiaoquan Yu, Michael Cawte, Yasmine Sfendla, Andreas Sadawsky, Andrew Doherty, Prahlad Warszawski and Matthew Woolley.\\



\newcommand{\beginsupplement}{
        \setcounter{table}{0}
        \renewcommand{\thetable}{S\arabic{table}}
        \setcounter{figure}{0}
        \renewcommand{\thefigure}{S\arabic{figure}}
        \renewcommand{\thesection}{S\arabic{section}}
        }

\title{Coherent vortex dynamics in a strongly-interacting superfluid on a silicon chip: \\
Supplementary Materials}

\author{Yauhen P. Sachkou, Christopher G. Baker, Glen I. Harris, \\
Oliver R. Stockdale,  Stefan Forstner, Matthew T. Reeves, Xin He,  \\
David L. McAuslan, Ashton S. Bradley, Matthew J. Davis, \& \\
Warwick P. Bowen}

%
%

\baselineskip24pt

\maketitle


\beginsupplement

\setcounter{section}{0}”

\vspace{-20 mm}

\onehalfspacing

\section{Experimental details}

\label{Sec:Experimental_details}

Our experimental system consists of a whispering-gallery-mode (WGM) microtoroidal optical cavity placed inside a vacuum- and superfluid-tight sample chamber at the base of a closed-cycle $^3$He cryostat. The sample chamber is filled with $^4$He gas, with a pressure $P=70$ mTorr at temperature $T=2.9$ K. Cooling down across the superfluid phase transition temperature condenses the $^4$He gas into a few-nm-thick superfluid film, which coats the inside of the sample chamber and the optical microcavity \cite{harris_laser_2016_supp}. For the $^4$He gas pressure used in the experiment (70 mTorr) the superfluid phase transition temperature is around 1 K.  The experiment is carried out at $T=500$ mK, with a 7.5 nm superfluid film thickness.
More experimental details are available in reference \cite{harris_laser_2016_supp} and its supplementary information. The microcavity is fabricated from a wafer made of a 2 $\mu$m thick thermal oxide layer grown atop a silicon substrate. The high-quality thermally grown oxide ensures that both top and bottom surfaces of the fabricated resonator have typically $<1$ nm RMS roughness \cite{lee_chemically_2012}, leading to reduced vortex pinning (see section \ref{Sec:vortex_pinning_on_surfaces}). The silica microtoroid is isolated from the silicon substrate atop a silicon pedestal \cite{harris_laser_2016_supp}. Light from a NKT Photonics (Koheras Adjustik) fibre laser with wavelength $\lambda=1555.065$ nm is evanescently coupled into a resonator WGM (linewidth $\kappa/2\pi\approx22$ MHz) via a tapered fibre. Superfluid film motion modulates the effective optical path length for the light confined to the periphery of the microcavity \cite{harris_laser_2016_supp, baker_theoretical_2016_supp}. This modulation manifests as fluctuations on the optical phase which are resolved via balanced homodyne detection (New Focus 1817) implemented within a fiber interferometer. The photocurrent is recorded with an Agilent Technologies MSO7104A oscilloscope with a sampling rate of 2 MHz. 

\subsection{Monitoring of the vortex-induced frequency splitting}

In order to display the frequency splitting evolution data shown in Figure 3 of the main text, we record the balanced-detector photocurrent with an oscilloscope. The acquired data consists of 6 consecutive traces, separated by short gaps corresponding to the data saving time.  Each trace is 50 s long and contains 100 million points (sampling rate 2 MHz). Each 50 s trace is then broken down into 125 bins, each 0.4 s long. We Fourier-transform each bin in order to obtain $6\times125=750$ third-sound power spectra, as shown in Fig. \ref{Figthirdsoundmodeidentification}(A). We fit each tracked sound mode with a double-peaked Lorentzian function in order to acquire the frequency separation between the split peaks, see Fig. \ref{Figthirdsoundmodeidentification}(C). This allows us to plot the frequency splitting for the five tracked sound modes as a function of time, with 750 time steps during the 360 s experimental decay process.

\subsection{Geometric contribution to sound splitting}
\label{subsec:geo_splitting}

Small departures from circular symmetry of the resonator, for instance induced by the CO$_2$ laser reflow process used to form the microtoroid\cite{harris_laser_2016_supp}, may split the third sound modes even in the absence of vortices in the superfluid film. This geometric splitting~\cite{ellis_observation_1989_supp, forstner_modelling_2019_supp} of the sound modes is analogous to the roughness-induced optical doublet splitting well known in high-Q optical microcavities.

In our experiment, most sound modes exhibit some small ($< 1$ kHz) degree of geometric splitting\cite{harris_laser_2016_supp}. This splitting is easily distinguishable from vortex-induced splitting, as it is of smaller magnitude, constant in time, and can be determined by observing the native splitting present when  repeatedly cycling through the superfluid transition temperature. To account for it, we remove its contribution to the total splitting $\Delta f_{\mathrm{total}}$ in order to isolate the vortex-induced contribution $\Delta f_{\mathrm{vortex}}$ using the
following relationship \cite{ellis_observation_1989_supp}:
\begin{equation}
\Delta f_{\mathrm{vortex}}=\sqrt{\Delta f_{\mathrm{total}}^2-\Delta f_{\mathrm{geo}}^2}
\label{Equationgeometricsplitting}
\end{equation}
The data presented in Figure 3 of the main text corresponds to the vortex-induced splitting, with the geometric splitting contribution removed.

\section{Third-sound mode identification}
\label{Sec:modes_identification}

\begin{figure}
\centering
\includegraphics[width=\textwidth]{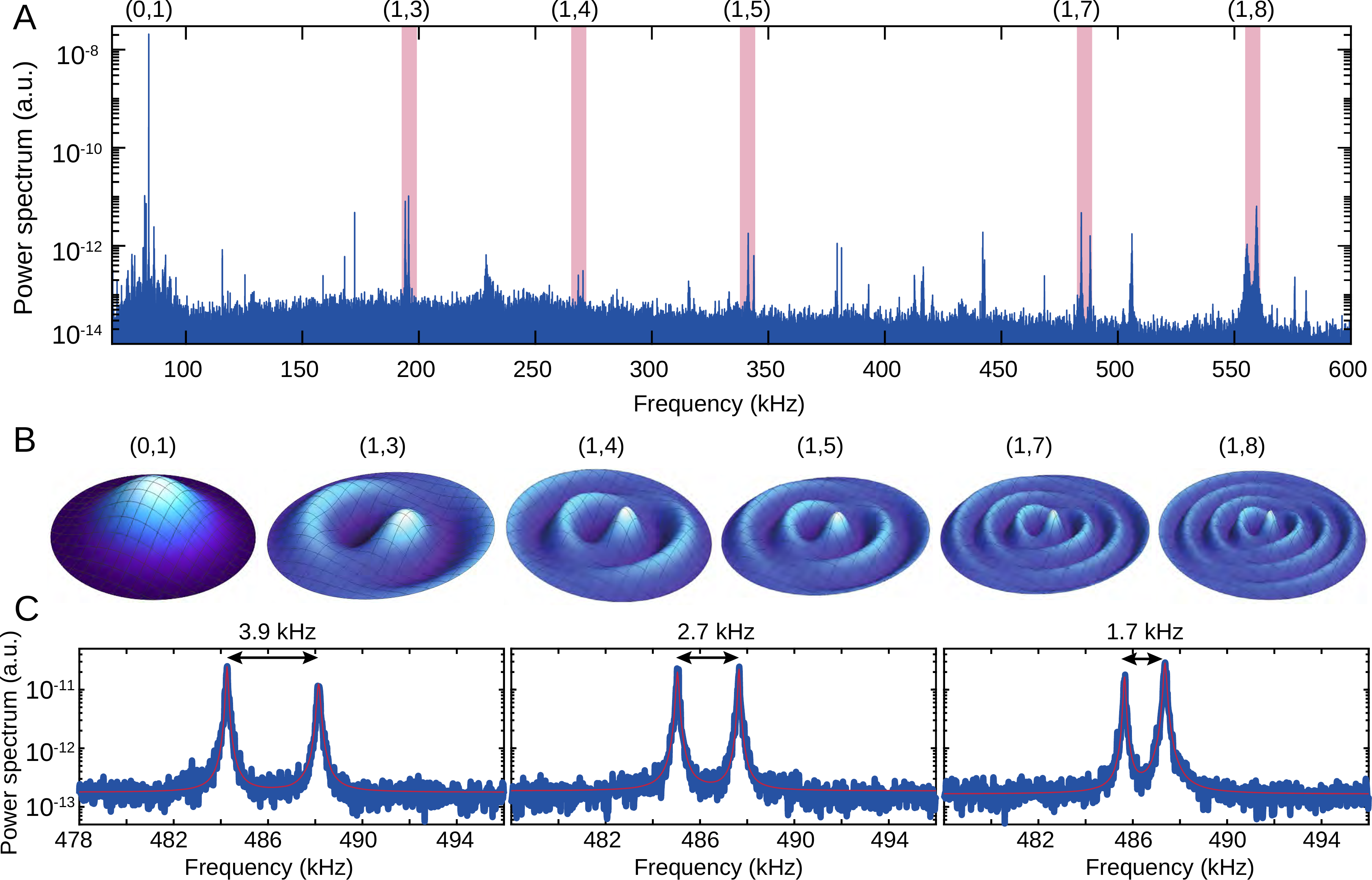}
\caption{(A) Third-sound power spectrum. The five sound modes monitored in the experiments are highlighted in red. The red bars are positioned at the theoretically predicted sound eigenfrequencies of a  30 $\mu$m radius disk resonator with free boundary conditions and a 7.5 nm film thickness \cite{baker_theoretical_2016_supp}, in good agreement with the experimental frequencies. Note that mode (1,6) is not used in the experiments, as its proximity to another sound mode precludes reliable peak fitting.  
(B) Surface deformation profiles of the 6 sound modes labelled in (A). (C) Example of the mode splitting decay observed in the experiments. The three panels show, from left to right, the splitting decay of mode (1,7) as the experiment progresses. Red line: double Lorentzian peak-fit to the data.}
\label{Figthirdsoundmodeidentification}
\end{figure}

Figure \ref{Figthirdsoundmodeidentification}(A) shows a power spectrum of the microcavity transmitted light, revealing the presence of multiple superfluid third-sound modes.
Correct identification of the sound modes is important in order to ascertain how each sound mode is differently affected by the presence of vortices \cite{forstner_modelling_2019_supp}.
As we discuss in section \ref{sectioninfluenceofpedestalonbesselmodes} below, the presence of the pedestal has a negligible effect on the sound eigenfrequencies, such that the experimental spectrum can be fitted with the eigenfrequencies of a disk resonator.  These are Bessel modes of the first kind, identified by their ($m$,$n$) mode numbers, which respectively correspond to the mode's azimuthal and radial order (see section \ref{sectionequationsofeigenmodesdiskannulus}).

We find that our experimental spectrum can be well reproduced  by the eigenfrequencies of a 30 $\mu$m radius disk resonator with free boundary conditions and a 7.5 nm film thickness\footnote{The film thickness (via its influence on the speed of sound) is used as the sole fitting parameter to match the Bessel mode spectrum. The fitted value of film thickness (7.5 nm) is in good agreement with values extracted from the optical cavity frequency shift \cite{harris_laser_2016_supp}.}, with typical frequency discrepancies on the order of a percent \cite{baker_theoretical_2016_supp}.  Note that the frequency spacing of higher-order Bessel modes is not harmonic, which allows us to discriminate between free and fixed boundary conditions \cite{baker_theoretical_2016_supp}.

Rotationally invariant third-sound modes ($m=0$) are most efficient at modulating the effective  optical path length of the resonator \cite{baker_photoelastic_2014,baker_theoretical_2016_supp}, and thus generally have the highest optomechanical coupling rate and hence the largest signal-to-noise ratio in the power spectrum. This is indeed what we observe, with the fundamental `drumhead' (0,1) Bessel mode having the largest signal-to-noise ratio in the experiments, see Fig. \ref{Figthirdsoundmodeidentification}(A).  Rotationally invariant modes, however, cannot be decomposed on the basis of clockwise and counterclockwise propagating modes. As such, we do not expect these modes to display any vortex-induced splitting \cite{forstner_modelling_2019_supp}. Again, this is what we observe, with the (0,1) mode remaining unperturbed by the vortex generation process, while  the (1,3), (1,4), (1,5), (1,7) and (1,8) modes all exhibit splitting, as seen in Fig. \ref{Figthirdsoundmodeidentification}(A).

\subsection{Influence of pedestal on third-sound modes}
\label{sectioninfluenceofpedestalonbesselmodes}

Here we consider the influence of the silica microtoroid's silicon pedestal on the superfluid helium sound eigenmodes on the underside of the toroid (\ref{sectionmechanicalfrequencyshiftduetopedestal}), as well as the pedestal's influence on the splitting experienced by these sound waves (\ref{sectionsplittingchangeduetopedestal}).

\subsubsection{Mechanical frequency}
\label{sectionmechanicalfrequencyshiftduetopedestal}

Using the Finite Element Method (FEM) techniques outlined in reference \cite{forstner_modelling_2019_supp}, we compare the sound eigenfrequencies on an annular domain with free-free boundary conditions at the edge of the disk as well at the level of the pedestal, to the eigenfrequencies on a simple disk geometry of same outer radius.  The results are summarized in Table \ref{Tablecomparisondiskannulus}. The effect of the pedestal on the third sound frequencies is negligible, typically less than one percent. This is because the presence of the pedestal does not significantly alter the sound eigenmode shapes, as shown for three different eigenmodes in Figure \ref{Figeffectofpedestal} and discussed in \ref{sectionequationsofeigenmodesdiskannulus}.

\subsubsection{Mode frequency splitting}
\label{sectionsplittingchangeduetopedestal}

\begin{table}
\begin{tabular}{|c|c|c|c|c|c|c|}

\hline 
Mode & $f_{\mathrm{disk}}$ (Hz)  & $f_\mathrm{annulus}$ (Hz) & $\Delta f_{\mathrm{disk}}$ (Hz) & $\Delta f_{\mathrm{annulus}}$ (Hz) & $f_{\mathrm{error}}$ (\%) & $\Delta f_{\mathrm{error}}$ (\%) \\ 
\hline 
\hline
(0,1) & 57318 & 57426 & - & - & 0.19 & - \\ 
\hline 
(1,6) & 269484 & 267301 & 193.3 & 188.3 & 0.81 & 2.6 \\ 
\hline
(1,8) & 363633 & 360127 & 261.8 & 256.7 & 0.96 & 1.95 \\ 
\hline 
\end{tabular} 
\caption{Comparison table for the eigenfrequencies and splitting in a disk vs annular geometry. Frequency and splitting values quoted for the eigenmodes of a 30 micron radius disk / a 30 micron outer radius annulus (see eigenmodes in Figure \ref{Figeffectofpedestalonfreqandsplitting}). Frequency values are quoted for a 10 nm thick superfluid film. The frequency error ($f_{\mathrm{error}}$) and splitting error ($\Delta f_{\mathrm{error}}$) are typically quite small, below 3 \%. Modes with $m=0$ cannot be decomposed in the basis of clockwise and counter-clockwise rotating modes and experience therefore no splitting.}
\label{Tablecomparisondiskannulus}
\end{table}

\begin{figure}
\centering
\includegraphics[width=0.9\textwidth]{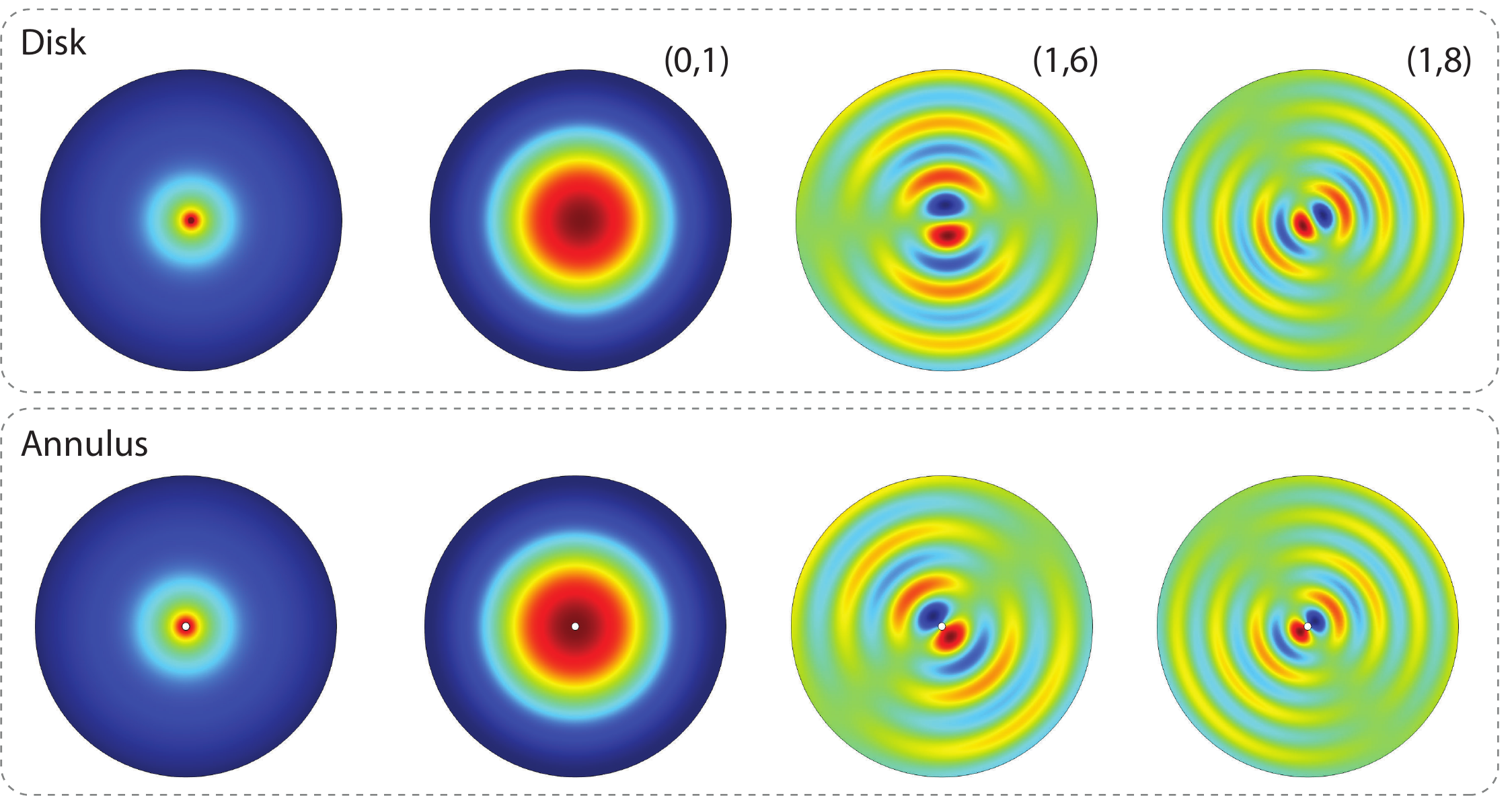}
\caption{ Leftmost image for both disk and annulus: calculated flow field due to quantized circulation around a centered point-vortex (disk) or a persistent current around a pedestal with a 1.5 $\mu$m diameter (annulus). Rightmost images: Out-of-plane displacement profile for three distinct sound eigenmodes. Color code: Red=positive; blue=negative; green=no displacement. The ($m$,$n$) numbers respectively correspond to the mode's azimuthal and radial order, see Eqs. (\ref{EqBesseldisk}) and (\ref{EqBesselannulus}) in \ref{sectionequationsofeigenmodesdiskannulus}. }
\label{Figeffectofpedestalonfreqandsplitting}
\end{figure}

Similarly, we compare the sound mode frequency splitting $\Delta f$ due to a vortex at the center of a disk geometry to the splitting due to a persistent current around the pedestal in an annular domain.  The results are also summarized in Table \ref{Tablecomparisondiskannulus}. These values are computed using the finite element simulation techniques outlined in reference \cite{forstner_modelling_2019_supp}. The splitting differences are again small, typically below 3\%. For this reason, it is reasonable to neglect the presence of the pedestal and model the underside of the microtoroid as a circular disk resonator, so far as the mechanical eigenmodes and the splitting they experience is concerned. (The presence of the pedestal is taken into account in the kinetic energy calculations discussed in section \ref{Sec:pinning_potential}).

\subsubsection{Eigenmodes of a disk and annulus}
\label{sectionequationsofeigenmodesdiskannulus}

Here we briefly describe the analytical expression of the superfluid third sound eigenmodes.
The eigenmodes on a disk are given by \cite{baker_theoretical_2016_supp}:
\begin{equation}
\eta_{m,n} \left( r, \theta \right)=\eta_0\,J_m\left(\zeta_{m,n} \frac{r}{R} \right) \cos \left( m \theta \right),
\label{EqBesseldisk}
\end{equation}
where $\eta_{m,n}$ describes the out-of-plane deformation of the superfluid surface for the $(m, n)$ mode as a function of polar coordinates $r$ and $\theta$. The ($m$,$n$) numbers respectively correspond to the mode's azimuthal and radial order, and $\zeta_{m,n}$ is a frequency parameter depending on the mode order and the boundary conditions.
The eigenmodes on an annulus \cite{baker_theoretical_2016_supp} are given by:
\begin{equation}
\eta_{m,n} \left( r, \theta \right)=\eta_0 \left(J_m\left(\zeta_{m,n} \frac{r}{R}\right) + \alpha Y_m\left(\zeta_{m,n} \frac{r}{R} \right) \right) \cos \left( m \theta \right)
\label{EqBesselannulus}
\end{equation}
Where $J_m$ and $Y_m$ are respectively Bessel functions of the first and second kind of order $m$. 
For small values $r_p \ll R$, as is the case in our experiments, $\alpha\ll1$  as $Y_m$ diverges at $r=0$, and the eigenmodes of the annulus are very close to those of the disk.

\section{Dissipative Point-Vortex Dynamics}
\label{sectionpointvortexmodel}

Here we introduce the dissipative Point Vortex Model (PVM) discussed in the main text. We employ this model in order to:
\begin{itemize}
\item Calculate the metastable quasi-equilibrium states of the system (sections \ref{subsec:metastable_state} and \ref{sectionexpandingvortexclustersbecomeflattop}).
\item Provide a fit to our experimentally measured vortex decay dynamics and extract the dissipation factor $\gamma$ and diffusion coefficient $D$ (section \ref{subsec:implementation_of_PVM}).
\end{itemize}

\subsection{Model}

The motion of quantum vortices of circulation $\kappa_j = s_j\, h/m$ (with $s_j \in \mathbb{Z}$ the vortex charge)  in a thin superfluid film of density $\rho_s$, thickness $d$, and temperature $T$ can be described by the equation of motion~\cite{ambegaokar_dissipation_1978,ambegaokar_dynamics_1980}
\begin{align}
\dv{\vb{r}_i}{t} = \vb{v}_s^{i} + C(\vb{v}_n - \vb{v}_s^{i} )  + s_i  \left (\frac{D h \rho_s d  }{m k_B T}\right) \vu{z}\times (\vb{v}_n - \vb{v}_s^{i}). \label{EOM}
\end{align}
Here $\vb{v}_n$ is the velocity of the normal fluid and $\vb{v}_s^i$ is the local superfluid velocity, evaluated at $\mathbf{r}_i$, excluding the self-divergent velocity field of the vortex at $\mathbf{r}_i$  \cite{ambegaokar_dynamics_1980}. The parameters $C$ and $D$ are phenomenological mutual friction coefficients, which originate from the scattering of quasiparticles such as phonons, rotons, and ripplons by the vortex cores; these parameters are dependent on temperature, film thickness, and the complex interactions of vortices with defects in the substrate~\cite{ambegaokar_dissipation_1978,ambegaokar_dynamics_1980}. Note that in the absence of friction, a vortex simply moves with the local superfluid velocity, as in the case of ordinary point-vortex dynamics of an ideal fluid~\cite{newton_n-vortex_2013}. The diffusion coefficient $D$ is typically $\lesssim \hbar /m$ near the Berezinskii-Kosterlitz-Thouless transition~\cite{ambegaokar_dynamics_1980,huber_vortex_1980}, and decays algebraically below $T_{BKT}$~\cite{adams_vortex_1987_supp}.  The constant $C$ ranges between 0 and 1.

In our system the normal fluid is viscously clamped to the surface of the microtoroid~\cite{harris_laser_2016_supp} and we may therefore assume $\vb{v}_n = \vb{0}$, giving 
\begin{align}
\dv{\vb{r}_i}{t} = (1-C)\vb{v}_s^{i}  - s_i \, \gamma\, (\vu{z}\times  \vb{v}_s^{i}). \label{simplifiedEOM}
\end{align}
where we have defined the dimensionless dissipation coefficient $\gamma = \left (D h \rho_s d  /m k_B T\right)$.
The superfluid velocity $\mathbf{v}_s$ generated by the vortices is determined by the Green's function of the domain~\cite{ambegaokar_dynamics_1980}. Here we approximate the  surface of the microtoroid as a hard-walled circular disk of radius $R$. The velocity at vortex $i$  is thus given in terms of the relative vortex positions as
\begin{align}
\mathbf{v}_s^{i} = \frac{1}{2\pi}
\sum_{j\neq i }\frac{\kappa_j}{r_{ij}^2}
\begin{pmatrix}
-y_{ij}\\
x_{ij}\\
\end{pmatrix} 
+ \frac{1}{2\pi} \sum_j\frac{\bar{\kappa}_j}{\bar{r}_{ij}^2}
\begin{pmatrix}
-\bar{y}_{ij}\\
\bar{x}_{ij}\\
\end{pmatrix},\label{eqnOfMotion}
\end{align}
where $x_{ij} = x_i-x_j$ and, $r_{ij}^2 = x_{ij}^2 + y_{ij}^2$. The barred terms $\bar{x}_{ij} = x_i-\bar{x}_j$, etc.  correspond to image vortices with $\bar \kappa_j = -\kappa_j$ placed outside the disk at the inverse point  $\vb{\bar{r}}_i = {R^2\vb{r}_i}/{|\vb{r}_i|^2}$. These fictitious image vortices enforce the boundary condition $\mathbf{v}_s \cdot \hat{\mathbf{n}} |_{r=R} = 0$, i.e., the flow across the boundary normal $\hat{\mathbf{n}}$ is zero at the boundary $r=R$.

The kinetic energy of the fluid can be expressed in terms of the relative vortex positions. In a disk of radius $R$, the vortex Hamiltonian for $N$ vortices with core size $a_0$, is given by~\cite{buhler_statistical_2002,marchand_vortex_2006}
\begin{align}
H = &- \frac{\rho_sd}{4\pi}\sum_{i<j}^{N,N}\kappa_i\kappa_j\ln\qty(\frac{r_{ij}^2}{Ra_0}) + \frac{\rho_sd}{4\pi}\sum_{i = 1}^N\kappa_i^2\ln\qty(\frac{R^2-r_i^2}{Ra_0})\nonumber  \\ &+ \frac{\rho_sd}{4\pi}\sum_{i<j}^{N,N}\kappa_i\kappa_j\ln\qty(\frac{R^4 - 2R^2\vb{r}_i\cdot\vb{r}_j+r_i^2r_j^2}{R^3a_0}).
\end{align}
In the absence of friction $(\gamma= C = 0)$,  $H$ generates the vortex dynamics, described in \eqn{eqnOfMotion}, from Hamilton's equations as
\begin{align}
\kappa_i\dv{x_i}{t} &= \pdv{H}{y_i}, & \kappa_i\dv{y_i}{t} = -\pdv{H}{x_i}. \label{hamil}
\end{align}

The parameter $C$ is non-dissipative (it does not influence the relative distances between vortices), and therefore does not contribute to the decay of the superflow~\cite{ambegaokar_dissipation_1978}. We therefore set $C=0$. As shall be addressed later in section \ref{subsec:implementation_of_PVM}, we measure the dissipation factor to be on the order of $\gamma\sim10^{-6}$. As the dimensionless parameters $\gamma$ and $C$ originate from similar microscopic processes, we expect that they would be of comparable magnitude, as is indeed the case in bulk superfluid helium~\cite{hanninen_vortex_2014,barenghi_experimental_2014,donnelly_observed_1998}; since $\gamma \ll 1$, the assumption $(1-C) \approx 1$ is therefore a reasonable approximation.

We add two phenomenological features to the model by hand: 
\begin{itemize}
\item Pinning of vortices to the pedestal of the microtoroid. In our experiments, we infer a large multi-quantum vortex is pinned to toroid's pedestal. To model this, we assume rigid pinning, forcing \eqn{simplifiedEOM} to equal zero for the vortices pinned at the origin.  
\item Vortex-antivortex annihilation. Vortex-antivortex pairs in a superfluid annihilate once they approach each other within a distance comparable to the core size~\cite{jones_motions_1982}, which is not included in Eqs. \ref{hamil}. To account for the annihilation of free vortices in the orbiting cluster with quantized circulation of the opposite sign trapped around the pedestal, we remove a free vortex from the simulation when it reaches the radius of the pedestal, $r_p = R/30$, along with one quantum of circulation from the pedestal.
\end{itemize}

\subsection{Metastable States}
\label{subsec:metastable_state}

\begin{figure}
\centering
\includegraphics[width=10cm]{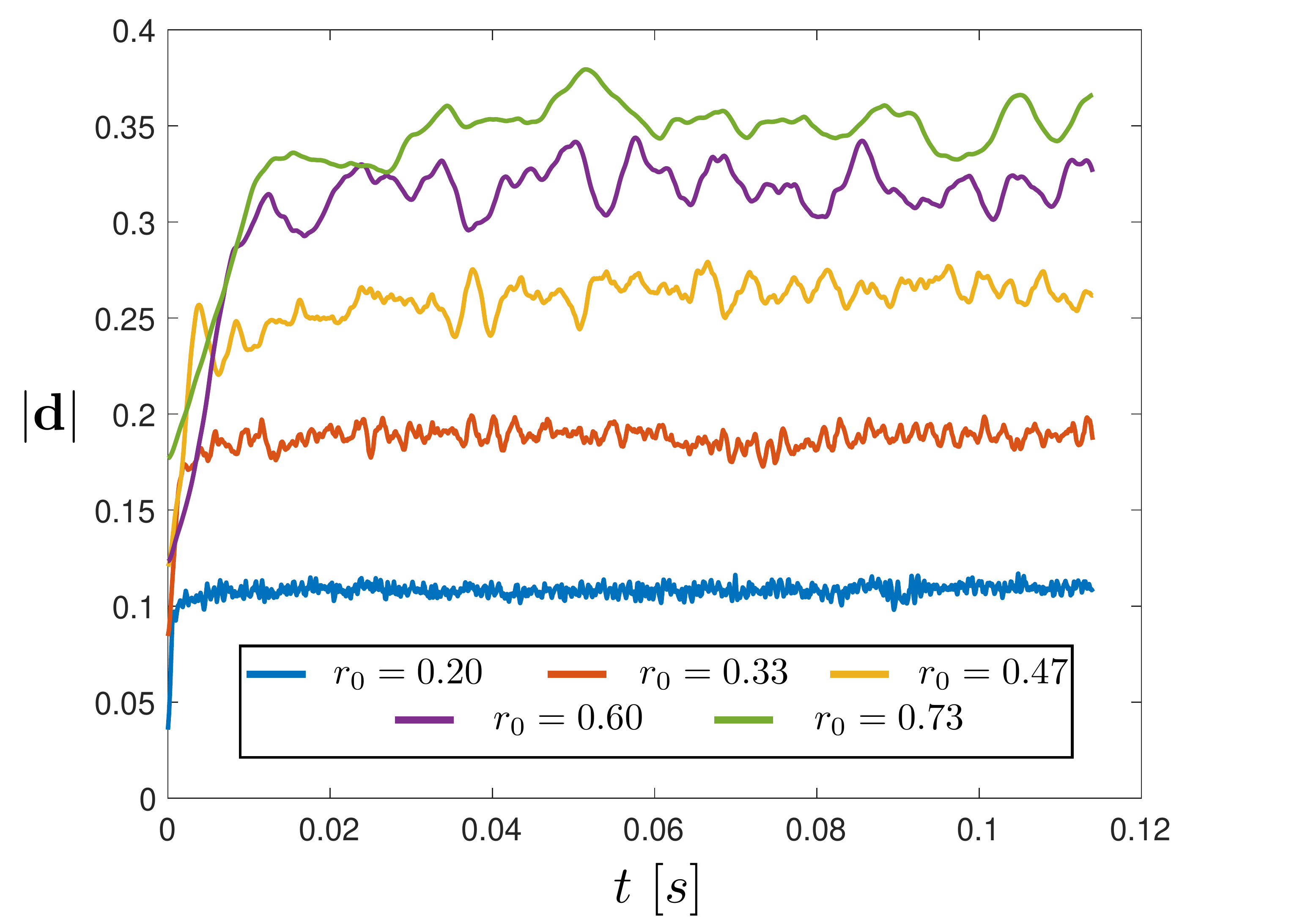}
\caption{Dipole moment of $N=17$ vortex cluster, for vortices initialized in a ring formation of  varying initial radius $r_0$. Each curve represents an average of 50 PVM simulation runs. The ring initialization radius $r_0$ is given here as a fraction of the disk radius R.}
\label{dipPlot}
\end{figure}

As discussed in the main text, we find that a cluster of negative vortices that orbit the pinned multi-quantum vortex quickly forms a metastable state. To show this, we evolve a cluster over time without dissipation and find that the dipole moment, defined as
\begin{equation}
d \equiv |\vb{d}| =\qty|\frac{1}{N}\sum_i^N s_i\, \vb{r}_i|,
\end{equation}
quickly becomes a non-zero constant and persists indefinitely. We run the system with zero dissipation as the time scale of the coherent dynamics is approximately six orders of magnitude smaller than the time scale of the dissipative dynamics. Hence we assume that the dissipative dynamics do not affect the cluster's tendency to form the metastable state. In \fig{dipPlot}, we plot the dipole moment for a cluster of 17 negative vortices, initialised in a ring of radius $r_0$, in the presence of a flow generated by a pinned $\kappa = 17 \, h/m$ vortex at the centre of the disk. 
Figure \ref{dipPlot} shows that the dipole moment of the largest ring reaches a finite value within $\tau \sim 20$ ms. The initial state of the experimental system after initialization (dipole separation $\sim$7 um), determined by fitting the frequency splitting of all sound modes, is most closely described by the blue curve in \fig{dipPlot}. As can be seen here, the metastable state is reached within $\tau \sim 5$ ms, a timescale much shorter than the observed decay of the splitting in the experiments (minutes). This value is also consistent with the estimation based on the $\tau\sim r_c^2 /N\kappa$ formula discussed in the main text.

As the system is in a microcanonical ensemble, only the energy, angular momentum, and vortex number of the initial distribution define the metastable state. However, it has been shown that stirring superfluid helium in an annular geometry leads to a ring of vortices \cite{fetter_low-lying_1967_supp}. As such, we reduce here the parameter space that describes the metastable state to just kinetic energy, $K$, and vortex number, $N$.

With this, we calculate the metastable states over the entire parameter space of $(N,K)$. We range the vortex number between $N=2$ to $N=19$ (as $N=1$ corresponds to the trivial case where the vortex simply orbits at a constant radius). For each cluster number, we vary the radius of the ring from $r_p$ through $R=28.50\ \mu \text{m}$ in steps of $\Delta R \simeq 0.1532\ \mu \text{m}$ (which directly correlates to incremental changes in kinetic energy), giving $n=180$ metastable states for each vortex number.

\begin{figure}
\centering
\includegraphics[width=13cm]{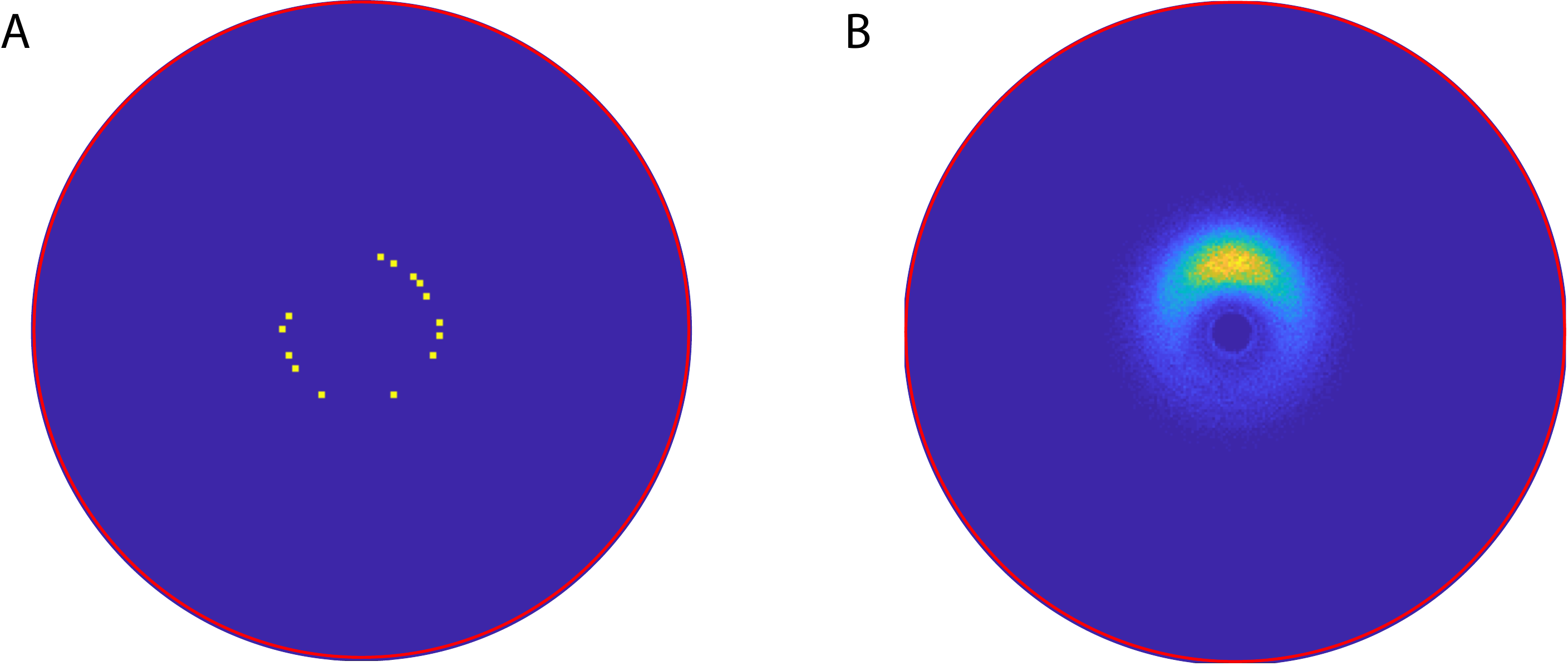}
\caption{(A) Sample initial ring distribution where $N=16$. Yellow dots correspond to the free  vortices in the system, which orbit around the pinned circulation in the center of the resonator (not shown here). Red circle: resonator outer boundary. (B) Time-averaged metastable state over the entire simulation. At each time step, the vortex cluster has been rotated such that the dipole moment lies along the y-axis.}
\label{exampleMetastable}
\end{figure}

We characterize the metastable state by calculating  the mean density of vortex positions over the course of the simulation. As the vortices perform many orbits during a simulation, plotting this in a fixed reference frame would produce a doughnut-shaped rotationally invariant-density profile. In order to visualize the angular spread of the orbiting cluster, we instead go into a frame rotating with the vortex cluster: at each time step, we rotate the cluster such that its dipole moment lies along the y-axis. An example of this visualisation can be seen in Fig. \ref{exampleMetastable}(B).

\subsection{Implementation}
\label{subsec:implementation_of_PVM}

We solve the system of equations in \eqn{simplifiedEOM} to calculate the motion of vortices with energy damping. We initialize the positions of the 17 negative vortices via a random uniform distribution to form a ring, with the condition that the distribution must have a kinetic energy comparable to the initial energy in the experiment ($\sim 1\ \mathrm{aJ}$). We simulate the dynamics of the system thirty separate times with different initial vortex positions, each constrained by initial energy, with dissipation $\gamma = 0.03$. In Fig. 4 C\&D of the main text, a single simulation result is plotted (solid blue line), with the light-blue shading representing one standard deviation on either side of the mean in the thirty runs.

\begin{figure}
\centering
\includegraphics[width=\columnwidth]{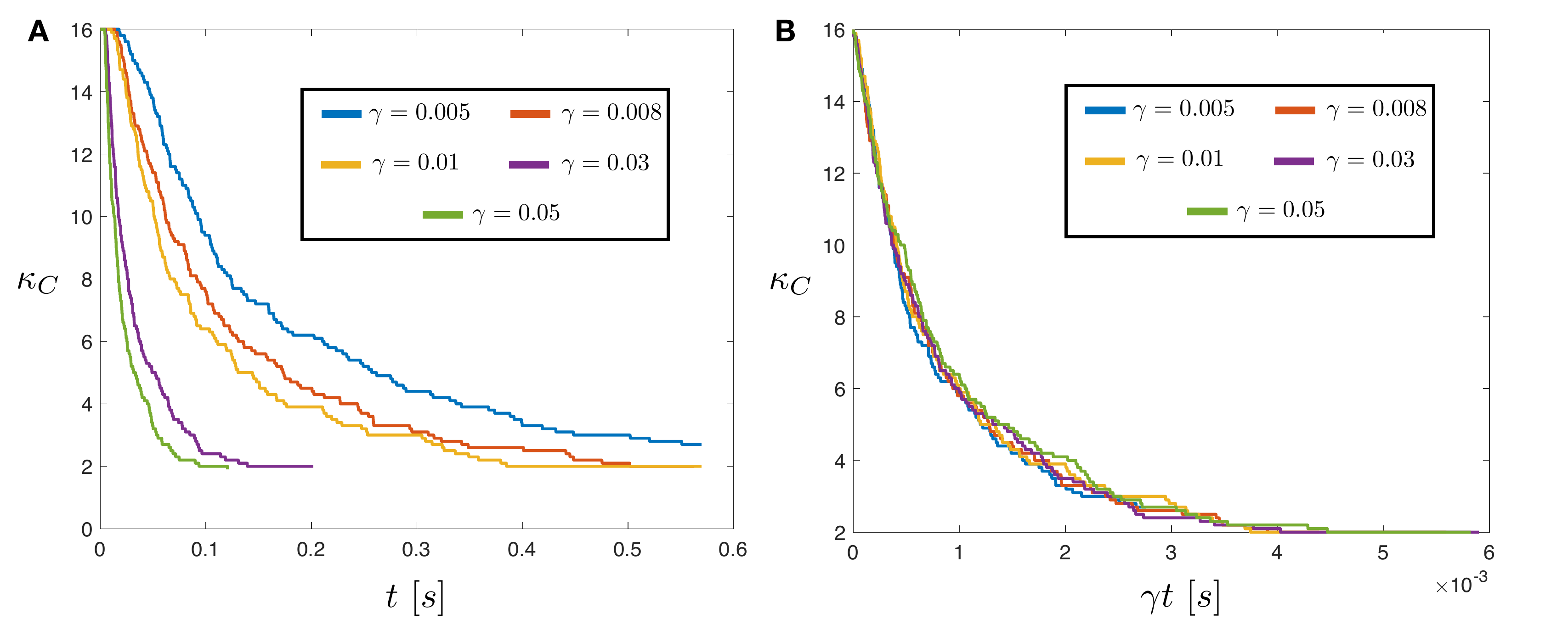}
\caption{(A) Decay of the charge of the pinned multi-quantum vortex ($\kappa_C$) for the different dissipation constants given in the legend. Each curve represents the average of 10 runs with different initial conditions constrained by the energy of the cluster. (B) Decay of the charge of the pinned multi-quantum vortex ($\kappa_C$) as in (A), but here the time axis has been scaled by the dimensionless dissipation constant $\gamma$. All curves fall on top of each other, highlighting the scaling of the dynamics with $\gamma$.}
\label{scaling1}
\end{figure}

Since the free vortex cluster evolves slowly through quasi-equilibrium states, we find that the macroscopic dynamics scale with the dissipation constant $\gamma$, as shown in \fig{scaling1}. Each curve seen in \fig{scaling1} is an average over 10 runs with different initial conditions. We expect that scaling will improve at lower dissipation rates as the time scale of energy loss will be far smaller than the time scale of coherent dynamics. As such, we choose a value of dissipation that is larger than the experimental one, and scale the results accordingly to best fit the experimental data, rescaling $\gamma$ in the process. 
This is necessary due to increasingly large simulation times for decreasing $\gamma$. We calculate the motion of vortices in time units of $mR^2/\hbar$. Within our simulations, we effectively run the system for approximately $\sim 50$ ms. After scaling our results, we find the experiment is consistent with a dissipation factor of $\gamma \sim 2\times 10^{-6}$, and hence a diffusion coefficient of approximately $D \sim 1\times10^{-16}\ \mathrm{m^2\ s^{-1}}$.

\subsection{Expanding Vortex Clusters}
\label{sectionexpandingvortexclustersbecomeflattop}

\begin{figure}
\centering
\includegraphics[width=15cm]{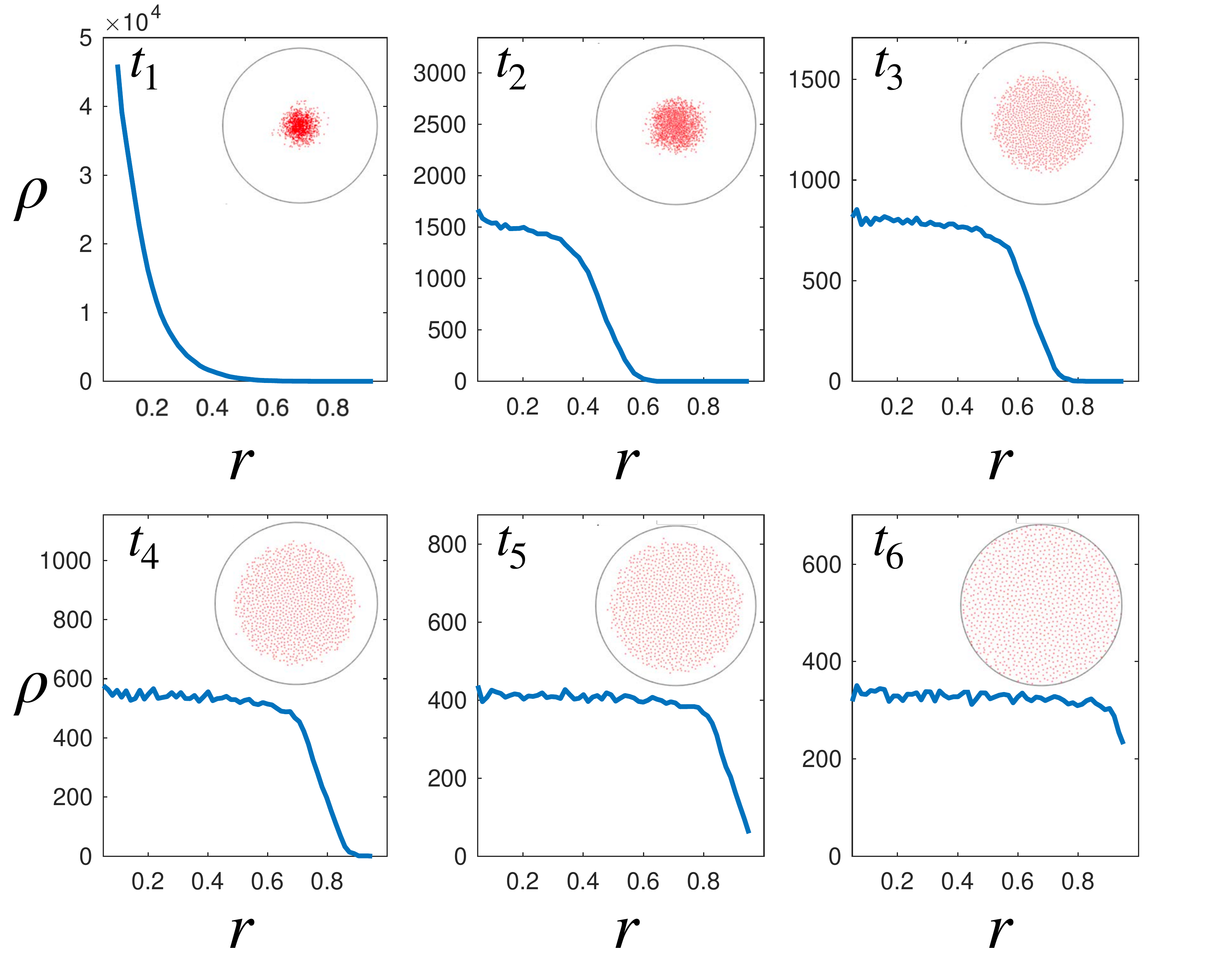}
\caption{Vortex cluster ($N=1000$) expansion from an initial Gaussian density distribution. Time progresses from left to right, top to bottom as per the labels. Despite the non-uniform initial distribution, vortices arrange themselves in such a way that the cluster rotates as a rigid body, resulting in an approximately uniform density at later times. This is supported by the insets in each figure, which show the physical position of vortices in the disc tending towards a uniform density as the cluster begins to crystallize.}
\label{densPlot}
\end{figure}

We use the point vortex model described above to observe the evolution of the density distribution of a vortex cluster initialised by a strongly non-uniform distribution. This simulation is used in the following section (\ref{Sec:vortex_decay_models}) to test a possible vortex decay scenario. In \fig{densPlot} we show that the density of a distribution of $N=1000$ vortices very quickly becomes uniform across the disc despite an initial Gaussian density distribution, indicated in the first panel of \fig{densPlot}. We have performed similar simulations with  initial density distributions corresponding to a ring or a flat-top, and observed the same rapid evolution towards uniform density.

\begin{figure}
\centering
\includegraphics[width=15cm]{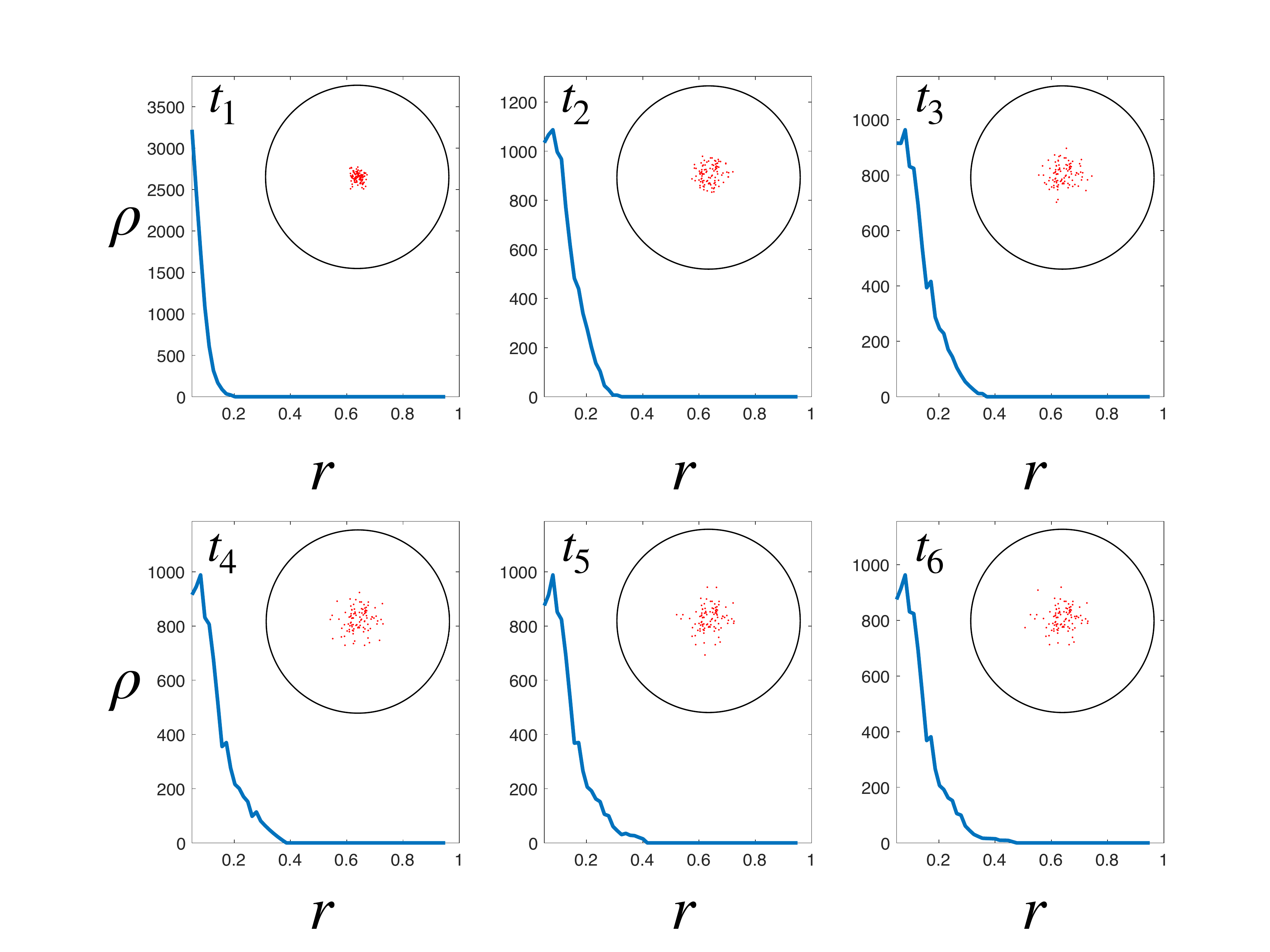}
\caption{Vortex cluster ($N=100$) expansion from an initial Gaussian density distribution with vortex pinning. Time progresses from left to right, top to bottom as per the labels. The pinning qualitatively changes the dynamical evolution of the cluster compared to the no-pinning case (Fig. \ref{densPlot}). The Gaussian density profile remains approximately Gaussian throughout the simulation, with the standard deviation of the distribution gradually increasing due to damping. Note that by $t_4$, the cluster is essentially no longer expanding. The choice of the mean-free-path $l$ sets here the maximal cluster width before the vortices freeze out.} 
\label{pinningDensity}
\end{figure}

We next simulate the expansion of a vortex cluster under the influence of pinning sites. These simulations are used in the next section to test an alternate vortex decay scenario. As previously, we begin with a Gaussian density distribution of vortices, but phenomenologically include vortex pinning within the point-vortex model. We model vortex pinning by asserting a mean-free-path, $\ell$, associated to the likelihood of a single vortex becoming pinned. The probability of a vortex `surviving' is assumed to be exponentially decreasing, and hence the probability of a vortex becoming pinned is $P(\text{pin}) = 1-\exp(-\Delta x/\ell)$, where $\Delta x$ is the distance travelled by the vortex\footnote{This simple model assumes an infinite pinning potential, such that once pinned, a vortex is permanently immobilized. Other simulations with a critical pinning velocity below which a vortex gets pinned, which allow for multiple pinning and unpinning events for each vortex, yield qualitatively similar results.}. 
Figure \ref{pinningDensity} shows that the density of a cluster of $N=100$ vortices remains approximately Gaussian for this situation, but become more spread out over the disc due to energy dissipation.

\section{Test of various decay models}
\label{Sec:vortex_decay_models}

\begin{figure}
\centering
\includegraphics[width=\textwidth]{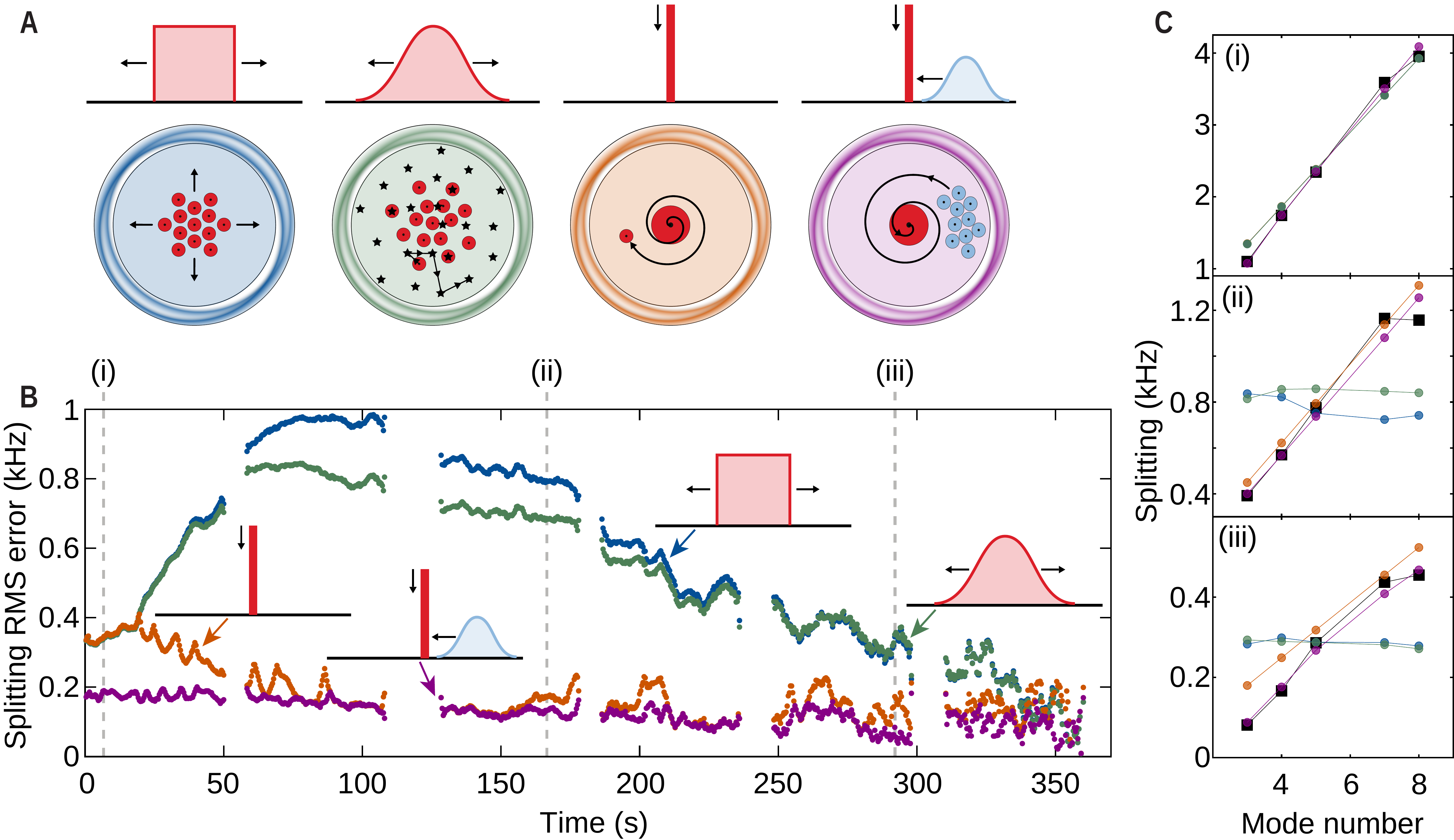}
\caption{A) Four different vortex decay scenarios. From left to right: flat-top (uniform density) expansion of a vortex cluster (blue disk); Gaussian expansion of a vortex cluster with pinning sites (green disk); decay of a macroscopic circulation through shedding of same-sign vortices (orange disk); decay of a macroscopic vortex dipole consisting of a pinned macroscopic circulation (red) and a cluster of free opposite-sign vortices (purple disk). The associated probability density is plotted above each scenario. (B) Calculated RMS error of each decay scenario across the 360 s experimental duration. The trace colors keep the same color scheme as in (A). At longer times, all scenarios converge towards a similar error as the experimental splitting has almost entirely decayed and our approach loses its discriminating power. (C) Experimentally-measured splitting of the five tracked sound modes ((1,3), (1,4), (1,5), (1,7) and (1,8)) (black squares) and simulated splitting corresponding to the four decay scenarios listed above. Each scenario is identified by the same color code as in (A). Each sound mode is identified by its azimuthal mode number $m$, \textit{e.g.} `8' corresponds to mode (1,8). Plots \textit{i}, \textit{ii} and \textit{iii} correspond to the times indicated by the dashed grey lines in (B). }
\label{Figvariousdecaymodels}
\end{figure}

Because the frequency splitting experienced by each third sound mode due to the presence of a vortex is unique (see Figure 1 of the main text, and reference \cite{forstner_modelling_2019_supp}),
the ability to monitor the splitting on multiple sound modes simultaneously allows us to infer spatial information about the experimental vortex distribution. We show here how this capability enables us to discriminate between several different vortex decay scenarios. We test four different scenarios which may account for the experimentally observed decay of the frequency splitting (see sketches in Figure \ref{Figvariousdecaymodels}(A)):

\begin{itemize}

\item Scenario 1. The experimentally observed splitting is initially due to a tight cluster of same-signed vortices  located at the center of the resonator. Such a distribution could be initialized by a circular stirring of the superfluid film, resulting in a centripetal Magnus force on one sign of vortices. After initialization, dissipation in the system will drive a radial expansion of the tightly-packed cluster. As discussed in section \ref{sectionexpandingvortexclustersbecomeflattop}, point-vortex simulations reveal that a tight vortex cluster will rapidly relax into an expanding flat-top (\textit{i.e.} spatially uniform) distribution, irrespective of the initial cluster arrangement. This flat-top expansion scenario is depicted on the blue resonator in Figure \ref{Figvariousdecaymodels}(A).\\

\item  Scenario 2. The experimentally observed splitting is due to a tight cluster of same-signed vortices located in the center of the resonator, as in scenario 1. The difference here lies in the presence of pinning sites (black stars) on the resonator surface. 
Expansion of the tightly-packed cluster in the presence of pinning sites resembles diffusive Brownian motion whereby vortices hop from pinning site to pinning site with a slow outward drift due to dissipation (see individual vortex trajectory marked by black arrows). Under such conditions the spatial probability density distribution of the expanding cluster can be approximated by a Gaussian (see section \ref{sectionexpandingvortexclustersbecomeflattop} for further details and point-vortex simulations).   This Gaussian expansion scenario is depicted on the green resonator in Figure \ref{Figvariousdecaymodels}(A).\\

\item  Scenario 3.  The experimentally observed splitting is due to a macroscopic persistent current around the pedestal at the center of the resonator (see discussion of the pinning potential due to the pedestal in section \ref{Sec:pinning_potential}).
This persistent current then decays through the slow shedding of same-sign vortices. Due to dissipation, these vortices will rapidly spiral out of the resonator,
such that the probability density can be modelled by a temporally decaying delta function in the center of the resonator. Due to the strong potential energy barrier for the vortex shedding process (see section \ref{subsec:metastability_of_pinned_flow_around_pedestal}), such a scenario is unlikely but is nonetheless considered here for completeness. This persistent current decay scenario is depicted on the orange resonator in Figure \ref{Figvariousdecaymodels}(A).  \\

\item  Scenario 4.  The experimentally observed splitting is due to a macroscopic persistent current pinned by the pedestal at the center of the resonator, and a cluster of orbiting free vortices of opposite sign. Such a distribution can be initialized by a superfluid flow exceeding the critical velocity up the device pedestal, leading to the generation of pairs of opposite-sign vortices. One sign pins to the pedestal while the other forms into a freely orbiting cluster. After initialization, dissipation will cause the orbiting cluster to spiral inwards, where it will annihilate with the persistent current on the pedestal. This vortex dipole decay scenario is depicted on the purple resonator in Figure \ref{Figvariousdecaymodels}(A).\\
\end{itemize}

\begin{figure}
\centering
\includegraphics[width=.6\textwidth]{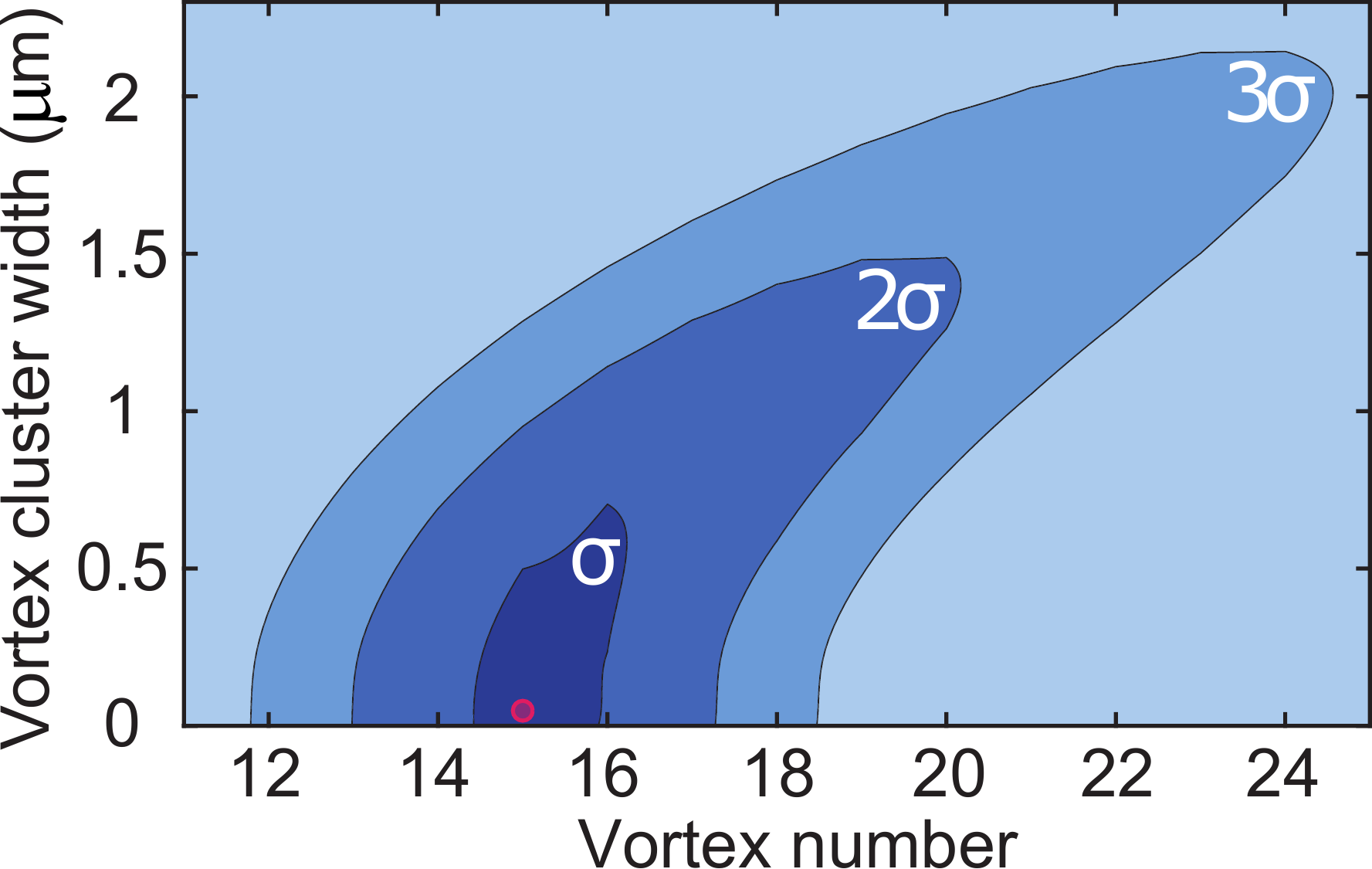}
\caption{RMS error map for the flat-top distribution decay scenario, computed at $t=0$. The labeled shaded regions correspond to the one-, two- and three-$\sigma$ standard deviation confidence intervals. The best fit to experiment at t=0 is provided by 15 vortices near the disk origin in this scenario (red circle).  }
\label{Figrmsclusermap}
\end{figure}

Each decay model is tested by comparing its predicted splitting on all sound modes to the experimental data. The results of this comparison are shown in Figure \ref{Figvariousdecaymodels} (B). 
We discuss here the flat-top decay model as an example of how the RMS error is calculated. The flat-top distribution is uniquely characterized by an initial vortex number and distribution radius. Looking at the initial splitting of the five tracked sound modes, we find the combination of vortex number and distribution radius which most closely matches the experimental results, and the associated confidence intervals, as illustrated in Figure \ref{Figrmsclusermap}.  This initial vortex number is then kept constant throughout the fitting procedure, as the distribution expands with no annihilation events. At all subsequent experimental time steps, the optimal radius which minimizes the RMS error is inferred. The fitting procedure thus produces an estimate of the distribution radius versus time, as well as an RMS error of the fitting procedure versus time, as plotted in the blue trace in Figure \ref{Figvariousdecaymodels}(B).  The calculation of the RMS error for the other decay scenarios follows a similar approach.
In scenario 3, the optimization is performed over $N$, the number of circulation quanta around the pedestal. In the vortex dipole decay case (purple), since the vortex number is also no longer constant due to annihilation events, the optimization at every time step occurs over a two-dimensional space (vortex number and kinetic energy), as in the initial fitting step of the flat-top distribution shown in Fig. \ref{Figrmsclusermap}.

All scenarios start with a tightly-packed initial vortex distribution or macroscopic circulation  at the center of the resonator. This is both because we expect the stirring process to generate vortices at the top of the pedestal, where the superflow is fastest, and because the initial ratio of the measured experimental splittings $\Delta f_{\mathrm{exp}}$ closely matches the  theoretical splitting $\Delta f_{\mathrm{th}}$ ratios expected from vortices positioned in the center of the resonator:
\begin{equation}
\frac{\Delta f_{\mathrm{exp}, \,m}\left( t=0 \right)}{\Delta f_{\mathrm{exp},m'}\left( t=0 \right)} \simeq \frac{\Delta f_{\mathrm{th}, \,m}\left( r=0 \right)}{\Delta f_{\mathrm{th},m'}\left( r=0 \right)}
\end{equation}
Here the subscripts $m$ and $m'$ refer to different sound mode azimuthal orders, and $\Delta f_{\mathrm{th}}\left(r\right)$ refers to the theoretically calculated vortex-induced splitting functions, as shown in Figure 1 of the main text. These splitting functions are independently obtained through Finite Element Method (FEM) calculation of  the overlap between the  vortex and sound flow fields  \cite{forstner_modelling_2019_supp}.

We note that while all scenarios predict a decaying splitting of all sound modes, as observed in the experiments, their relative influence on different sound modes is quite significant. This is striking in Fig. \ref{Figvariousdecaymodels}C(\textit{ii}) \& \ref{Figvariousdecaymodels}C(\textit{iii}), which show the predicted splitting on all five sound modes at the times marked by the dashed grey lines in Fig. \ref{Figvariousdecaymodels}(B). We see that both expanding cluster scenarios (flat-top in blue and Gaussian in green) predict similar splitting on all sound modes ---on the order of 0.8 kHz in (\textit{ii}) and 0.3 kHz in (\textit{iii})--- as the cluster expands. This is in strong disagreement with the experimental data (black squares) for which the splitting is an increasing function of mode number. 
This mismatch, visible in the much larger RMS errors for these decay scenarios in Figure \ref{Figvariousdecaymodels}(B), illustrates how the ability to track the frequency splitting of multiple sound modes simultaneously enables us to infer spatial information on the experimental vortex distribution.
We observe that the vortex-dipole scenario discussed in the main text (purple) systematically provides the best agreement with the experimentally measured splitting throughout the entire decay process.

\section{Vortex pinning on surfaces }
\label{Sec:vortex_pinning_on_surfaces}

The presence of surface roughness on a substrate that is in contact with superfluid helium limits the ability of vortices to move freely. Indeed, in thin superfluid films, strong pinning sites can freeze out vortex motion altogether \cite{ellis_quantum_1993_supp}. For example, surface roughness is thought to be the primary cause of discrepancy between measurements of the vortex diffusivity around the BKT transition\cite{adams_vortex_1987_supp}. It is for this reason that studies of vortex interaction, dissipation and diffusion are critically dependent on surface quality. 

\begin{figure}
\centering
\includegraphics[width=\textwidth]{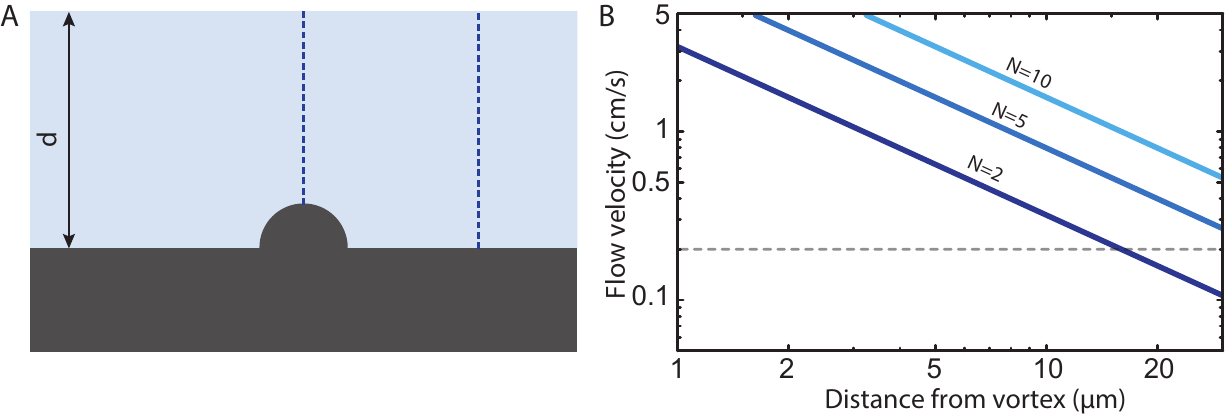}
\caption{(A) A simplified model of pinning due to a spherical defect in a thin film of superfluid helium of thickness $d$. The vortex located on the defect (central dashed line) has less kinetic energy than one off the defect (right), due to its reduced length.  
(B) Background flow velocity induced by $N$ circulation quanta trapped around the pedestal, versus radial offset from the disk center. The horizontal line indicates a flow velocity of $0.2\,\rm cm/s$.}
\label{Fig_pinning}
\end{figure}

The effect of surface roughness on vortex motion can be understood by considering the case of a superfluid film residing on a smooth substrate with a defect in the form of a spherical protuberance, as shown in Fig.~\ref{Fig_pinning}(A) ~\cite{browne_vortex_1982,schwarz_three-dimensional_1985}. From this simple picture, one can see that a vortex confined to the defect will be in an energetically favourable state. Indeed, for a constant film thickness d, a vortex located on the defect (center) will have less kinetic energy than one located off the defect (right) due to its shorter length. This energy difference is associated to a trapping potential, which naturally will grow with increasing defect size.
Similarly, from this simple picture one can see that for a given defect size, a thick film will lead to weaker pinning, as the height difference introduced by the defect becomes proportionately smaller.

Depinning occurs when the forces acting upon the vortex due to the background flow exceed the strength of the pinning potential. Clear signatures of depinning in superfluid films of thickness $3\,\rm nm$ have been reported, for example, in Ref.\cite{ellis_quantum_1993_supp}. In that experiment, the frequency splitting of degenerate third sound modes is slightly reduced when ``stirring'' at flow velocities above $150 \, \rm cm/s$, but below the superfluid critical velocity. The observed change in splitting corresponds to depinning of only a small fraction of the total number vortices, namely $~10^{2}-10^{3}$ compared to $10^{5}$, respectively. 
This observation suggests that those liberated vortices were attached to weak pinning sites, while the remaining vortices were pinned to much stronger pinning sites, which may be consistent with the comparatively larger surface roughness of their evaporated gold substrate.

Here, we ascribe the lack of any apparent contribution of pinning to the vortex dynamics to the combination of an atomically smooth surface and a relatively thick superfluid film. As discussed in sections \ref{sectionpointvortexmodel} \& \ref{Sec:vortex_decay_models} and Figure 4 of the main text, our data supports a model for vortex motion that is characterized by the decay of a vortex dipole down to a few remaining vortices after a period of minutes. As the vortex dipole decays, the background flow velocity across the disk decreases, as illustrated in Fig.~\ref{Fig_pinning}b. The observed decay is consistent with point-vortex modelling in the absence of pinning sites (see section \ref{sectionpointvortexmodel}), suggesting that pinning does not play a significant role in vortex dynamics in this system. Towards the end of the experimental run, the background flow induced by two vortices pinned to the pedestal creates flow velocities below $\sim 0.2\rm \, cm/s$ over the majority the disk surface. Indeed, the free vortex cluster explores these outer-parts of the disk because, as discussed in the main text, upon an annihilation event the cluster moves towards the periphery. These very low depinning velocities correspond to extremely small surface defects, likely to be of comparable size to the vortex core, consistent with the expected atomic-scale roughness of our commercially purchased silicon wafers.

\section{Pinning potential of the pedestal }
\label{Sec:pinning_potential}

\begin{figure}
\centering
\includegraphics[width=0.65\textwidth]{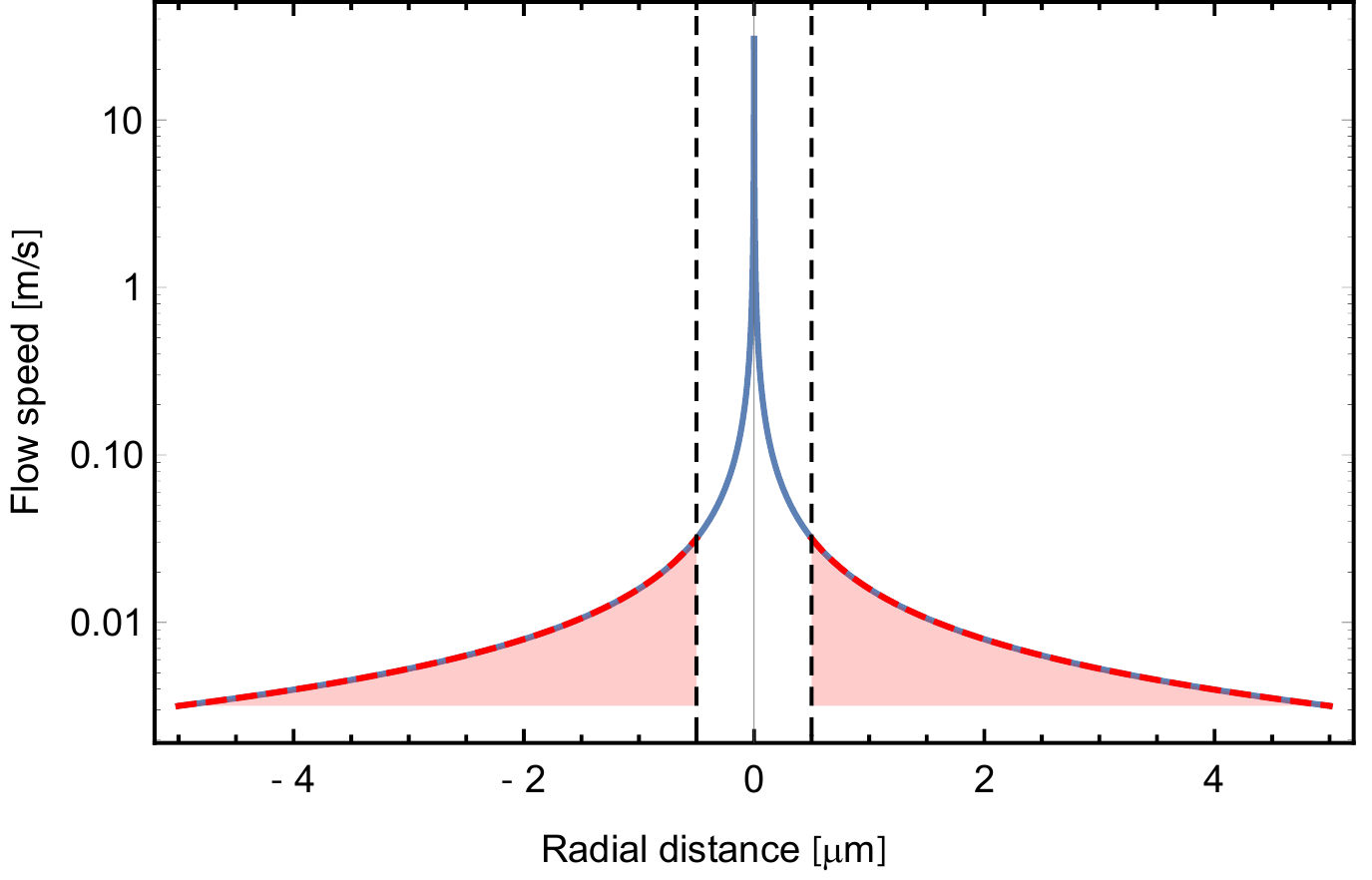}
\caption{Superfluid flow velocity as a function of the radial distance from the core, for a single circulation quantum $\kappa=h/m_4=9.98\times 10^{-8}$ m$^{-2}$s$^{-1}$ (blue line). By clipping the high-velocity component of the flow field, quantized circulation around the micron-sized microtoroid pedestal (materialized by vertical dashed gray lines), is energetically favourable. }
\label{Figeffectofpedestal}
\end{figure}

While superfluid flow is purely irrotational, a superfluid may contain vorticity consisting of 
irrotational flow around a topological defect in the film within which the vorticity is located  
\cite{tilley_superfluidity_1990_supp, donnelly_quantized_1991}. This topological defect may take the form of an Angstrom-sized normal fluid core in $^4$He, or circulation around a macroscopic barrier in a non-simply connected geometry. The kinetic energy $E_k$ associated with such a flow around a centred circular defect in a circular domain of radius R is given by:
\begin{equation}
E_{k,n}=\frac{1}{2}\,\rho_s \frac{\left(N\,\kappa\right)^2}{2\pi}\,\ln\left(\frac{R}{a_0}\right),
\label{Eqkineticenergycenteredvortexindisk}
\end{equation}
with $\rho_s$ the superfluid surface density, $a_0$ the defect radius and $N\times \kappa$ the quantized circulation around the defect, where $\kappa=h/m_{\mathrm{He}}=9.98\times 10^{-8}$ m$^{-2}$s$^{-1}$ is a single circulation quantum in $^4$He. Because circulation around a macroscopic barrier --such as the microtoroid pedestal ($a_0=r_p\sim 10^{-6}$ m) -- clips the high-velocity region of the flow (see Figure \ref{Figeffectofpedestal}), it is energetically favourable compared to circulation around a vortex core in the film  ($a_0\sim10^{-10}$ m). This has been observed in the stirring of superfluids in annular containers, where the vorticity preferentially takes the form of persistent flow around the inner annular boundary, and vortices appear in the fluid itself only for much larger rotation speeds \cite{donnelly_stability_1966, fetter_low-lying_1967_supp}.

For our $R=30$ micron resonator, with pedestal radius $r_p\sim 1$ $\mu$m, there is approximately three times less energy in the flow for quantized circulation around the pedestal than around a normal fluid core\footnote{The kinetic energy of a centred normal-fluid core vortex on a disk of radius $R=30\, \mu$m  is $\sim 1\times 10^{-20}$ J, which corresponds to $\sim$2000 $k_B\,T$ (for $T=500$ mK and a 10 nm thick film).}. However, placing increasingly large circulation around the pedestal becomes less advantageous as the energy cost of each additional circulation quantum $E_{k,N+1}-E_{k,N}$ is boosted  by $2\,N$, with $N$ the number of quanta already present, as shown in Equation (\ref{Eqkineticenergycenteredvortexindisk}). This thereby provides an upper bound on the number of circulation quanta around the pedestal before it becomes energetically favourable to shed these and form vortices in the fluid. In order to quantitatively estimate the metastability of the pinning around the pedestal, we numerically calculate the kinetic energy barrier for the shedding of vortices.

\subsection{Flow-field calculation}

\begin{figure}
\centering
\includegraphics[width=.8\textwidth]{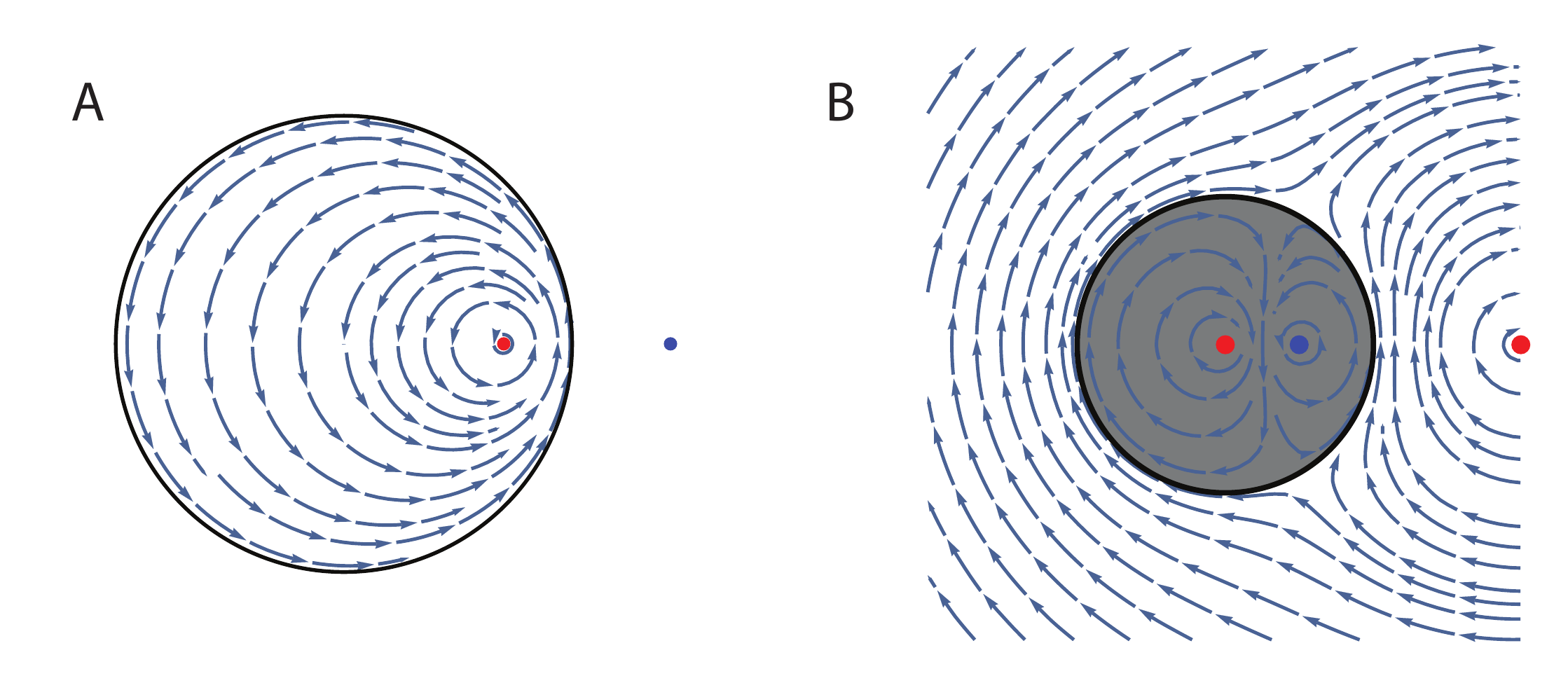}
\caption{(A) Flow streamlines of a point vortex (red dot) offset from the center inside a circular domain (black circle). The no-flow boundary condition is enforced by the presence of an opposite signed image-vortex outside the boundary (blue dot) \cite{lamb_hydrodynamics_1993}. (B) Flow streamlines of a point vortex (rightmost red dot) outside a cylindrical obstacle (gray disk) in a 2D plane. The no-flow boundary condition is enforced by an opposite signed image-vortex inside the obstacle (blue dot) and a same-signed image vortex in the center of the obstacle (left red dot).}
\label{image_vortices_fig}
\end{figure}

There are two mechanisms through which the large quantized circulation may decay. First, through the  escape of vortices of the same sign as the circulation.  Second, a negative vortex may be generated at the resonator boundary and migrate inward, annihilating circulation on the pedestal.  We estimate the change in kinetic energy of the flow for these two processes.

The streamfunction $\Psi_{\mathrm{free}}$ of a point-vortex on the underside of the toroid can be approximated in cartesian coordinates by \cite{lamb_hydrodynamics_1993,forstner_modelling_2019_supp,reeves_quantum_2017,ashbee_dynamics_2014}:
\begin{equation}
\begin{aligned}
\Psi_{\mathrm{free}} \simeq \, &\frac{\kappa}{2\pi} \, \ln \left(  \left(\sqrt{\left(x-X_1\right)^2+y^2}\right)-\ln\left(\sqrt{\left(x-X_2\right)^2+y^2}\right) \right. \\
&+ \left. \ln\left( \sqrt{x^2+y^2}\right)-\ln\left(\sqrt{\left(x-X_3\right)^2+y^2}\right) \right)
\label{EqPsifreevortex}
\end{aligned}
\end{equation}

\begin{figure}
\centering
\includegraphics[width=\textwidth]{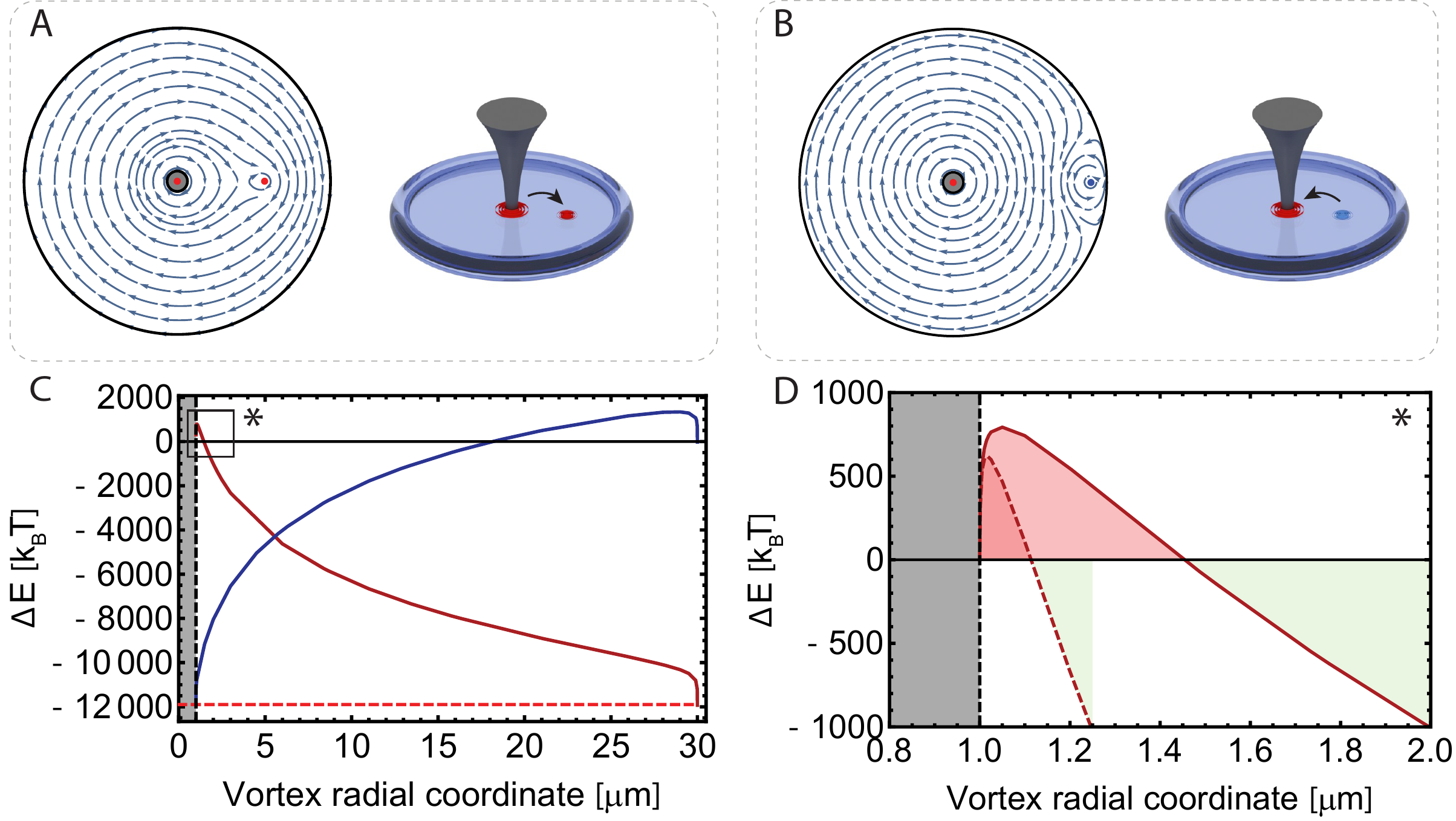}
\caption{Two alternate decay scenarios for a large ($\kappa \gg 1$) circulation trapped around the device pedestal. (A) Positive vortex (red) `tearing off' from the large positive pinned circulation around the central pedestal (greyed-out region). Blue arrows display the flow streamlines calculated from Eq. (\ref{EqvortexvelocityfromPsi}). (B) Incoming negative vortex (blue) annihilating with the positive circulation on the device pedestal. (C) Calculated kinetic energy cost of the decay processes illustrated in (A) and (B), starting from a positive circulation of $11\times\kappa$ around the pedestal. (red curve: escaping positive vortex; blue curve: incoming negative vortex). The dashed red line marks the theoretical energy drop from the removal of one circulation quantum. (D) Zoom-in of the region marked by an asterisk '*' in (C), highlighting the potential barrier for the shedding of positive vortices. Solid red line and dashed red line show the potential barriers starting from a positive circulation of $11\times\kappa$ and $31\times\kappa$ around the pedestal respectively. Greyed out region marks the location of the pedestal.}
\label{Figmetastabilityofthepinning}
\end{figure}

Here $X_1$ is the radial coordinate of the vortex (along the $x$ axis), $X_2=\frac{r_p^2}{X_1}$ is the radial coordinate of the opposite circulation image-vortex required to enforce no flow accross the pedestal boundary and $X_3=\frac{R^2}{X_1}$ is the radial coordinate of the opposite circulation image-vortex required to enforce no flow across the resonator boundary, as shown in Figure \ref{image_vortices_fig}. Because there are two boundaries in the problem, these image vortices also require their own image vortices, leading to an infinite series. However, for small values of $r_p \ll R$, as is the case in our experiments, the infinite series can safely be truncated after the first term, as in Eq. \ref{EqPsifreevortex}, while conserving the absence of flow through the resonator and pedestal boundaries, as visible in the calculated streamlines in Fig. \ref{Figmetastabilityofthepinning} (a) and (b). The streamfunction $\Psi_{\mathrm{pedestal}}$ of the  quantized circulation around the pedestal is given by:
\begin{equation}
\Psi_{\mathrm{pedestal}} = \frac{N\,\kappa}{2\pi} \, \ln\left( \sqrt{x^2+y^2}\right),
\label{EqPsicirculationpedestal}
\end{equation}
with $N$ the number of circulation quanta around the pedestal. The total streamfunction $\Psi_{\mathrm{tot}}$ describing the combined flow is then simply $\Psi_{\mathrm{tot}}=\Psi_{\mathrm{free}}+\Psi_{\mathrm{pedestal}}$.
From the streamfunction $\Psi_{\mathrm{tot}}$, the superfluid velocity components are given by:
\begin{equation}
v_{x}=\frac{\partial \Psi_{\mathrm{total}}}{\partial y};\qquad \mathrm{and}\qquad
v_{y}=-\frac{\partial \Psi_{\mathrm{total}}}{\partial x}
\label{EqvortexvelocityfromPsi}
\end{equation}
The kinetic energy of a given flow field is obtained through numerical integration of the calculated field (Eq. \ref{EqvortexvelocityfromPsi}) over the annular region contained within $r_p<r<R$.

\subsection{Metastability of the pinned flow around the pedestal}
\label{subsec:metastability_of_pinned_flow_around_pedestal}

Figure \ref{Figmetastabilityofthepinning}(A) and  \ref{Figmetastabilityofthepinning}(B) 
illustrate two alternate decay scenarios for a large circulation trapped around the device pedestal. First, a positive vortex (red) can be shed from the large positive pinned circulation around the central pedestal. Second, a  negative incoming vortex (blue) can  annihilate with the positive circulation on the device pedestal. The energy cost of both these scenarios is illustrated in Figure \ref{Figmetastabilityofthepinning}(C). The red curve plots the energy difference between the two following configurations:
\begin{itemize}
\item a positive circulation of 10$\kappa$ on the pedestal and a point vortex of circulation $\kappa$ at radial coordinate $x$
\item a positive circulation of 11$\kappa$ on the pedestal
\end{itemize}
If the point vortex can move far enough away from the pedestal, this escape process is energetically favourable ($\Delta E<0$), as the flow fields of the point vortex and the macroscopic circulation no longer sum up constructively. However, as shown in Fig. \ref{Figmetastabilityofthepinning}(D), there is an initial energy barrier on the order of several hundred $k_B T$ to do so. Indeed, in the initial stage of the vortex shedding process the flow fields of the pinned circulation and the point vortex still add up constructively, and do not offset the additional energy cost of the high velocity flow near the normal fluid core, as shown in Fig. \ref{Figeffectofpedestal}.

As discussed previously, the overall shape of the energy barrier will be dependent on the number of pinned circulation quanta $N$. For increasing circulation around the pedestal the barrier becomes narrower and shallower (dashed red line in Fig. \ref{Figmetastabilityofthepinning}(D)),  but the conclusion remains identical. While the lowest energy state of the system corresponds to no persistent current, a large energy barrier (on the order of several hundred $k_B \, T$ for the vortex numbers present in the experiment) explains why the circulation remains pinned around the device pedestal for the duration of the experiment.

Similarly, the creation of a negative vortex on the device boundary also incurs a large energy penalty due to the energy cost of the high-velocity region near the vortex core (see Fig. \ref{Figmetastabilityofthepinning}(C)). Note however that for radial coordinates $r<18$ $\mu$m, that energy cost is more than offset by the cancellation of the pinned persistent current flow field, such that the negative point-vortex introduces a negative kinetic energy to the system.


\end{document}